\documentclass[]{aastex63}

\shorttitle{HST/WFC3 Imaging Survey of NGC 6302}
\shortauthors{Kastner et al.}

\begin{document}

\title{Panchromatic HST/WFC3 Imaging Studies of Young, Rapidly Evolving Planetary Nebulae. \\ I. NGC 6302}

\correspondingauthor{Joel Kastner}
\email{jhk@cis.rit.edu}

%\author[ORCID]{Joel H. Kastner}
\author{Joel H. Kastner}
\affiliation{Center for Imaging Science 
  and Laboratory for Multiwavelength Astrophysics, 
  Rochester Institute of Technology, Rochester NY 14623, USA; jhk@cis.rit.edu}
\affiliation{School of Physics and Astronomy 
  and Laboratory for Multiwavelength Astrophysics, 
  Rochester Institute of Technology}

\author{Paula Moraga}
\affiliation{School of Physics and Astronomy 
  and Laboratory for Multiwavelength Astrophysics, 
  Rochester Institute of Technology}

\author{Bruce Balick}
\affiliation{Department of Astronomy, University of Washington, Seattle WA, 98195, USA}

\author{Jesse Bublitz}
\affiliation{School of Physics and Astronomy 
  and Laboratory for Multiwavelength Astrophysics, 
  Rochester Institute of Technology}

\author{Rodolfo Montez Jr.}
\affiliation{Center for Astrophysics | Harvard \& Smithsonian, Cambridge MA, 02138, USA}

\author{Adam Frank}
\affiliation{Department of Physics \& Astronomy, University of Rochester, Rochester NY, 14627, USA}

\author{Eric Blackman}
\affiliation{Department of Physics \& Astronomy, University of Rochester, Rochester NY, 14627, USA}

%\tableofcontents

% End of file `sample63.tex'.
\begin{abstract}
We present the results of a comprehensive, near-UV-to-near-IR Hubble Space Telescope WFC3 
imaging study of the young planetary nebula (PN) NGC 6302, the archetype of the class of extreme bi-lobed, pinched-waist PNe that are rich in dust and
molecular gas. The new WFC3 emission-line image suite 
clearly defines the dusty toroidal equatorial structure that bisects NGC 6302's polar lobes, and the fine structures
(clumps, knots, and filaments) within the lobes. The most striking aspect of the new WFC3 image suite is the bright, S-shaped 1.64 $\mu$m [Fe {\sc ii}] emission
that traces the southern interior of the east lobe rim 
and the northern interior of the west lobe rim, in point-symmetric
fashion. We interpret this [Fe~{\sc ii}] emitting region as a zone of shocks caused by ongoing, fast ($\sim$100 km s$^{-1}$), collimated, off-axis winds from NGC 6302's
central star(s). The [Fe~{\sc ii}] emission and a zone of dusty, N- and S-rich clumps near
the nebular symmetry axis form wedge-shaped structures on opposite sides of the core, with boundaries marked by sharp azimuthal ionization gradients. Comparison of our new 
images with earlier HST/WFC3 imaging reveals that the object previously
identified as NGC 6302's central star 
is a foreground field star. Shell-like inner lobe features may instead pinpoint the obscured central star's
actual position within the nebula's dusty central torus. The juxtaposition of
structures revealed in this HST/WFC3 imaging study of NGC 6302 presents a daunting challenge for models of the origin and evolution of bipolar PNe.  
\end{abstract}


\section{Introduction} \label{sec:intro}

{Planetary nebulae  (PNe) are the near-endpoints of stellar evolution for
intermediate-mass ($\sim$1--8 $M_\odot$) stars. Each PN provides a snapshot of
the brief ($\sim$$10^4$ yr duration) stage in which the outflowing, dusty
circumstellar envelope of an asymptotic
giant branch (AGB) star is ionized by its newly unveiled core, itself a future white dwarf \citep[e.g.,][]{Schoenberner1986}. 
PNe thereby offer our
last and arguably best look at the products of intermediate-mass stellar
nucleosynthesis, just before that material is incorporated into the
ISM \citep{Kwitter2014}. 

Though best known as photogenic, $\sim10^4$ K optical emission line
sources, a subset of PNe --- those with pinched-waist, bipolar
structures --- retain
significant masses of cold ($<100$ K), dense ($\sim10^4-10^6$ cm$^{-3}$) molecular
gas and dust that is irradiated, shocked, and continuously reshaped from within by UV and
winds from hot ($\sim$30--200 kK) central stars \citep[e.g.,][and
refs.\ therein]{Huggins1996,Kastner:1996ab,Bachiller1997,Huggins2005,Bublitz2019}. 
As the descendants of relatively massive progenitor stars
\citep{CorradiSchwarz1995}, many of these dusty, molecule-rich,
bipolar PNe are also the youngest and most morphologically and dynamically extreme objects
\citep{Kastner:1996ab, Balick:2002ab,Sahai:2011a}. 

An effective way to study PN-shaping stellar winds, and the rapid evolution of these winds, is
to observe their impacts on the slower but denser post-AGB material
downstream.  Classical PN shaping theory implicitly or explicitly posits that the driving force of
the active winds from the present-day central stars of PNe is isotropic radiative
momentum \citep{Kwok:1978ab}. However, the highly non-isotropic lobes of bipolar PNe must be formed
in some other manner, via the ram pressure in collimated winds or jets of
some sort \citep[e.g.,][]{Bujarrabal:2001lr,Akashi2018,Balick2019}. Evidence has steadily accumulated that
such collimated outflows are ultimately due to the influence of an (interacting) binary companion to the
mass-losing central star \citep[e.g.,][and references therein]{JonesBoffin2017,DeMarcoIzzard2017}, with
processes such as common envelope (CE)
evolution \citep[e.g.,][]{Livio1979,Soker:2004ao,GarciaSegura2018,Zou2020} or jets associated with a companion star's 
accretion disk \citep[e.g.,][]{Morris1987,Soker:1994ab,Chen2017} providing the mass launching and
collimation mechanisms. 

Narrow-band imaging with the {\it Hubble
Space Telescope} (HST) has long played a fundamental role in advancing
our understanding of these and other potential PN shaping mechanisms
\citep[e.g.,][]{Sahai:1998lr}. 
As HST neared the end of its third decade in operation, however, the community of PNe
researchers had yet to take full advantage of the potential of its
most capable imaging instrument, the Wide Field Camera 3
(WFC3). This changed in HST Cycle 27, when we used HST/WFC3 to
obtain the first comprehensive, contemporaneous sets of near-UV through near-IR (243 nm to 1.6 $\mu$m) emission-line
images of two especially structurally rich PNe, NGC 7027 and NGC 6302 \citep[see][for an initial overview of
these HST/WFC3 imaging surveys]{Kastner2020}. As we demonstrate in this paper, the broad, contemporaneous wavelength 
coverage of these WFC3 image suites yields full-nebula emission line image overlays and line ratio maps at $\sim$0.1" 
resolution that are free from distortions and artifacts caused by nebular proper motions or 
cross-instrument calibration uncertainties (among other potential pitfalls).

The subject of this paper, NGC 6302 (the Butterfly), is the archetype of the class of
extreme bi-lobed, pinched-waist PNe that are rich in dust and
molecular gas. As of this
writing, NGC 6302 had been featured or mentioned in more than 750 papers,
according to \verb+simbad+. Its bipolar lobes are very bright and rich in emission lines,
spanning a remarkably broad range of ionization states, from the far-UV through the mid-IR  \citep[e.g.,][]{Casassus2000,Feibelman2001,Molster2001,Groves2002}. Indeed, 
NGC 6302 has long served as an exemplar for studies of
high-excitation PNe \citep[e.g.,][]{Aller1981}, with estimates of its central star
temperature ranging from 220--250 kK \citep{Casassus2000,Wright2011} to as high as $\sim$400 kK
\citep{AshleyHyland1988}. Expansion
parallax distance estimates for NGC 6302 range from $\sim$0.81 kpc \citep{Lago2019} to $\sim$1.17 kpc
\citep[][]{Meaburn2008}, with a recent determination, adopted here, of 1.03$\pm$0.27 kpc \citep{GomezGordillo2020}. 
This proximity makes NGC 6302 a tempting subject for high-resolution
imaging studies aimed at understanding the origin and evolution of
bipolar structure in PNe. Indeed, prior to the comprehensive
Cycle 27 WFC3 survey described here, NGC 6302 had already been the subject of more
than two dozen HST images, including WFC3 imaging in several filters across the wavelength range
3700--6700 \AA\ \citep[e.g.,][]{Szyszka2009}. 

NGC 6302's central pinched waist is so dusty and its bipolar symmetry axis so highly inclined \citep[inclination of 75--78$^\circ$ with
respect to the line of sight;][]{Peretto2007,SantanderGarcia2017}, that, as dryly noted by
\citet{Aller1981}, ``[no] central star has been detected; the object
definitely is not a conventional planetary nebula.''  The dusty
central torus harbors ``mixed'' (C-rich and O-rich) chemistry,
displaying a combination of 
PAH features in the range $\sim$5--12 $\mu$m and crystalline silicate
and H$_2$O ice at longer IR wavelengths \citep{Molster2001,Kemper2002}. ALMA CO
mapping has revealed potential connections between this dense, dusty
molecular torus and the inner lobe regions near the waist of the nebula
\citep{SantanderGarcia2017}.  
%The CO velocity field across the torus suggests the presence of
%shocks accelerating the molecular gas and destroying CO present along the inner edge. 
The slowly ($\sim$8 km
s$^{-1}$) expanding equatorial molecular torus has a dynamical age of
$\sim$5000--7500 yr \citep{Peretto2007,SantanderGarcia2017}. In contrast, the large, 
open-ended bipolar lobes, which extend to $\sim$1.5 pc \citep[][]{Rao2018}, 
display very high expansion velocities \citep[$\sim$600 km s$^{-1}$;][]{Meaburn2005} and, hence, appear to have been ejected more recently 
and over a shorter ($\sim$2000 yr) timescale \citep{Meaburn2008,Szyszka2011}. 
Again quoting \citet{Aller1981}, ``The nebula is clearly the result of some explosive event; its whole
appearance suggests violent motions with some bilateral symmetry.'' 

Our new panchromatic HST/WFC3 imaging study of NGC 6302 is designed to elucidate
the structure and ionization patterns of the nebula on size scales from $\sim$100 au to $\sim$0.5 pc, so as to
constrain models of the history of its central star and the
evolution of its collimated outflows. In this paper, we
present the full suite of HST/WFC3 images of NGC 6302, as well as selected line ratio images, and we highlight key
results gleaned from these images thus far.  In forthcoming companion papers, we will provide an analogous treatment of the
HST/WFC3 imaging survey of NGC 7027 (Moraga et al., in prep.), and we will present an analysis of multi-epoch HST/WFC3 imaging of NGC 6302 
aimed at understanding its detailed kinematics (Balick et al., in prep.). 
The present paper is structured as follows: we describe the HST/WFC3 observations in
\S2; we present the resulting image suite, and briefly describe the most notable features of
the resulting image, in \S3; we present and analyze line ratio
images diagnostic of nebular extinction and excitation 
in \S4; we discuss the overall structure of the nebula's bipolar lobes, as revealed by the
HST/WFC3 image and line ratio suite, in \S5, while \S6 consists of a discussion of what our
images reveal (and can't reveal) about the central star; finally, in \S7, we
present a summary and the main conclusions gleaned from this HST/WFC3
imaging study of NGC 6302.

\section{Observations}

\begin{table}
\begin{center}
\caption{\sc HST/WFC3 Imaging Survey of NGC 6302:
  Observation Summary}
\label{tbl:summary}
\footnotesize
\begin{tabular}{ccccc}
\hline
Filter & $\lambda_0$ ($\Delta \lambda$)$^a$ &  line targeted & date & exp.\ \\
         & (nm) & & & (s) \\
\hline
\hline
FQ243N$^b$ & 246.8 (3.6) & [Ne {\sc iv}] $\lambda 2425$ & 2019-10-06 & 1110 \\
F343N  & 343.5 (25.0) & [Ne {\sc v}] $\lambda 3426$$^c$ & 2019-10-06 & 1110 \\
F487N & 487.1 (6.0) & H$\beta$ $\lambda 4861$ & 2019-10-06 & 1200 \\
F502N &  501.0 (6.5) & [O {\sc iii}] $\lambda\lambda 4959, 5007$ & 2020-03-13$^d$ & 1200\\
F656N &  656.1 (1.8) & H$\alpha$ $\lambda 6563$ & 2019-10-06 & 1200  \\
F658N & 658.4 (2.8) & [N {\sc ii}] $\lambda 6583$ & 2020-03-13 & 2600 \\
F673N &  676.6 (11.8) & [S {\sc ii}] $\lambda\lambda 6716, 6730$ & 2020-03-13$^d$ & 1290 \\
F110W & 1153.4 (443.0) & ``YJ band'' & 2019-10-05 & 556 \\
F128N & 1283.2 (15.9) & Pa$\beta$ 1.28 $\mu$m & 2019-10-05 & 506 \\
F130N & 1300.6 (15.6) & Pa$\beta$ continuum & 2019-10-05 & 506\\
F160W & 1536.9 (268.3) & ``H band'' & 2019-10-05 & 456 \\
F164N & 1640.4 (20.9) & [Fe {\sc ii}] 1.64 $\mu$m & 2019-10-05 & 1306 \\
F167N & 1664.2 (21.0) & [Fe {\sc ii}] continuum & 2019-10-05 & 1306 \\
\hline
\end{tabular}
\end{center}

{\sc Notes:} 
a) Filter pivot wavelength and bandwidth. b) Images in filter FQ243N
yielded poor signal, so are not presented here. c) Contamination from
continuum as well as O {\sc iii} $\lambda 3444$ at the $\sim$10\% percent level
\citep[][]{Groves2002}. d) Exposures with F502N and F673N in 2019 were
unusable due to gyro lock failures. 
\end{table}

The HST/WFC3 observations of NGC 6302 reported here were obtained during HST
Cycle 27 in 2019 October and 2020 March. The UVIS (CCD) and IR (HgCdTe array) imaging modules of WFC3 provide fields of view of
$\sim$$2.7'\times2.7'$ and $\sim$$2.27'\times2.27'$ at pixel scales of
0.04 arcsec/pixel and 0.13 arcsec/pixel, respectively.
Table~\ref{tbl:summary} provides a summary of the WFC3 filters used, lines targeted, observing
dates, and exposure times. The UVIS images were obtained in 2-point
GAP-LINE dither mode (DITHER-LINE for quad filter
FQ243N), and the IR images were obtained in 2-point DITHERBLOB
dither mode. Images obtained in 2019 October and  2020 March were
obtained at orientations of $-35.0^\circ$ and $140.2^\circ$,
respectively, as measured E of N with respect to the image Y axis.

Standard pipeline calibration and processing of the UVIS and IR
exposure sets, using {\tt CALWF3 v3.5.0},
was performed on all images\footnote{See
  https://hst-docs.stsci.edu/wfc3dhb/chapter-3-wfc3-data-calibration}. For
both UVIS and IR imaging, the calibrations
include bias correction, dark current subtraction,
flat field and shutter shading corrections, and cosmic-ray
rejection; pipeline UVIS image processing also includes CTE
correction. Following calibration, pipeline processing\footnote{https://hst-docs.stsci.edu/wfc3dhb/chapter-4-wfc3-images-distortion-correction-and-astrodrizzle} consisted of
geometric distortion corrections, World Coordinate System image
alignment, additional cosmic-ray rejection,  and merging of dithered
exposure sequences.  As a result of the simple two-point dither
observing mode we employed, cosmic-ray rejection was inefficient in the
inter-CCD ``gap'' region of the UVIS module, resulting in some
small-scale image artifacts along this region of the UVIS images.

We used the {\tt AstroDrizzle} software package\footnote{https://www.stsci.edu/scientific-community/software/drizzlepac.html}
to perform an additional round of post-pipeline processing of all
images, so as to (a) correct the registration of
dithered exposures that were poorly aligned in standard pipeline
processing (as was the case for most of the IR images), and (b) to refine relative image-to-image pointing across
filters, using common field stars as the positional references. We estimate that, after this post-pipeline processing, the relative 
pointing across all (UVIS and IR) images is accurate to
$\sim$0.5 pixel. 
Absolute pointing was further calibrated for a reference (IR)
image set using reference stars from
the Gaia Space Astrometry mission, by adapting standard  {\tt
  AstroDrizzle} scripts. 

The various emission
line and line ratio images
presented in this paper were generated from these astrometrically
calibrated images using appropriate
photometric calibrations and/or passband
responses, as obtained from the \verb+synphot+ software
package\footnote{https://www.stsci.edu/hst/instrumentation/reference-data-for-calibration-and-tools/synphot-throughput-tables.html}. Additional
small {\it ad hoc} pointing refinements were performed to generate difference images from the
F164N-F167N and F128N-F130N image pairs, so as to obtain
continuum-subtracted 1.28 $\mu$m
Pa$\beta$ and 1.64 $\mu$m [Fe {\sc ii}] images, respectively. These refinements
were verified via minimization of difference image residuals in
background sky regions and field star images. 

%\section{Results and Analysis}

\section{The HST/WFC3 image suite}

\begin{figure}[!ht]
\centering
\includegraphics[width=6in]{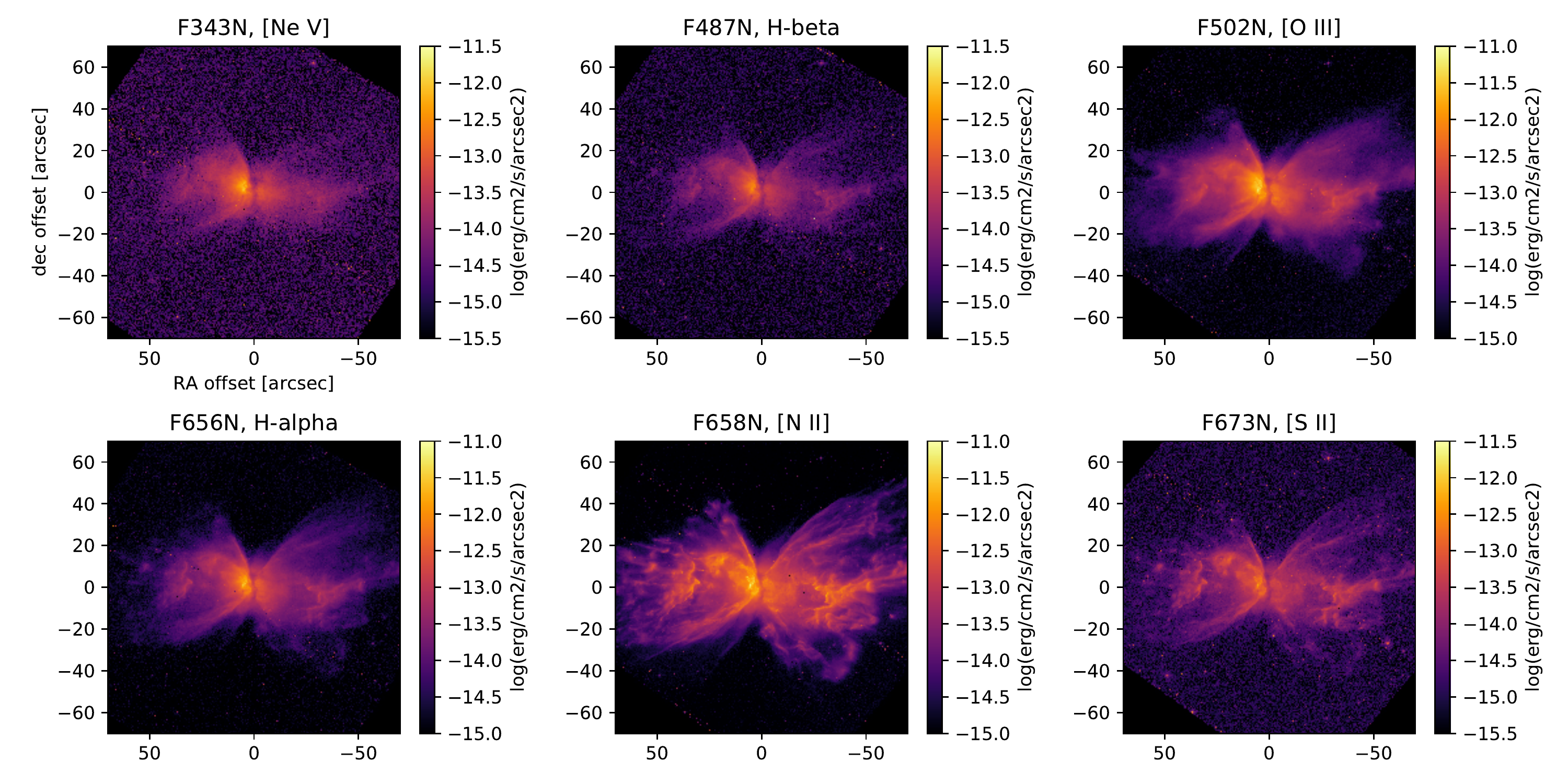}
\includegraphics[width=6in]{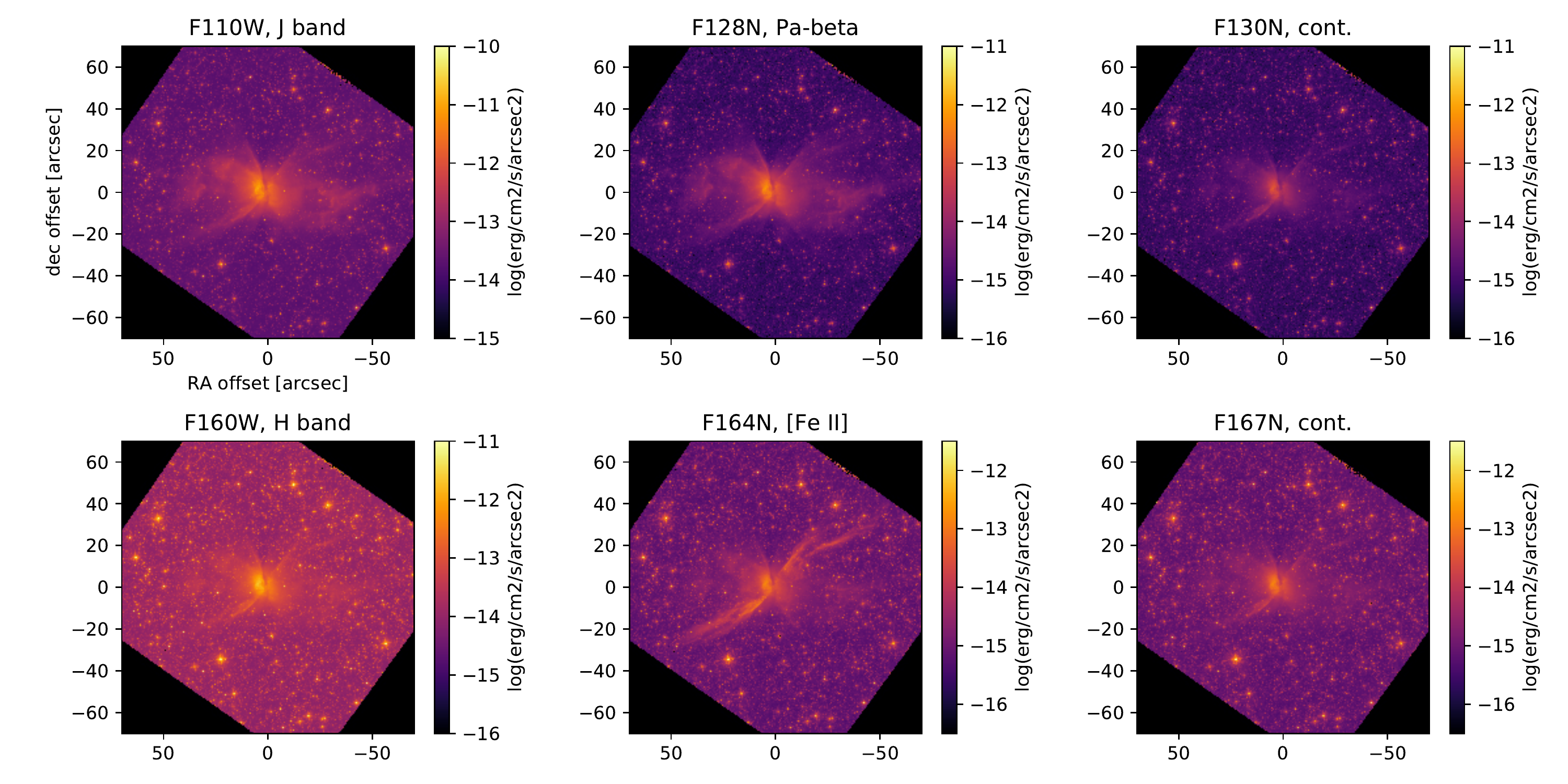}
\caption{
Suite of HST/WFC3 UVIS images obtained for NGC
6302. Field of view is $140'' \times 140''$; N is up and E is to the left.
}
\label{fig:NGC6302imageSuite}
\end{figure}

\begin{figure}[!ht]
\centering
\includegraphics[width=6in]{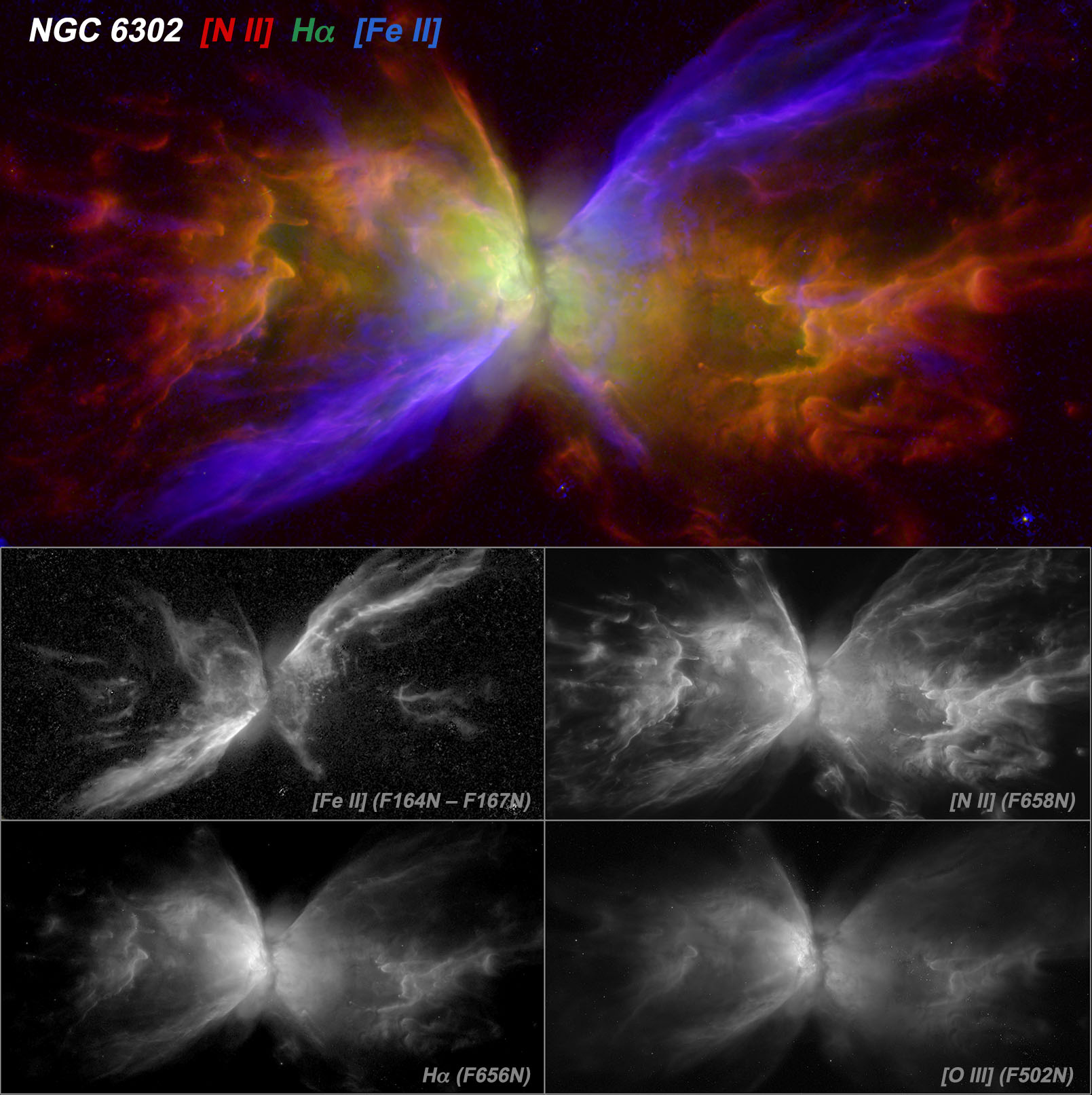}
\caption{Color overlay of WFC3 narrowband images of NGC 6302, with [Fe {\sc ii}]
(F164N$-$F167N difference image) coded blue, H$\alpha$ (F656N) coded green, and [N {\sc ii}]
(F658N) coded red. The three component images, along with the  [O {\sc iii}]
(F502N) image, are displayed in
greyscale in the lower panel. Field of view is $130'' \times 64''$; N is up and E is to the left.
}
\label{fig:NGC6302imageOverlay1}
\end{figure}

\begin{figure}[!ht]
\centering
\includegraphics[width=6in]{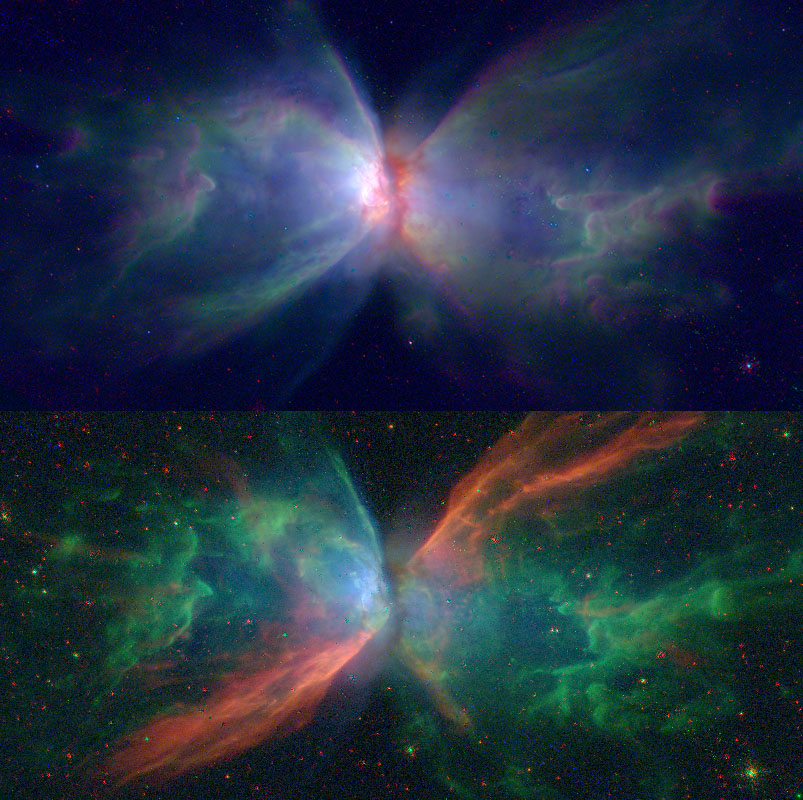}
\caption{Color overlays of WFC3 narrowband images of NGC 6302. 
Top: overlay with Pa$\beta$
(F128N$-$F130N difference image) coded red, [O {\sc iii}]
(F502N)  coded green, and [Ne {\sc v}]
(F343N) coded blue.
Fields of view are $130'' \times 64''$; N is up and E is to the left. Bottom:  overlay with [Fe {\sc ii}]
(F164N$-$F167N difference image) coded red, [S {\sc ii}]
(F673N)  coded green, and [Ne {\sc v}]
(F343N) coded blue. 
}
\label{fig:NGC6302imageOverlay2}
%\vspace{-.25in}
\end{figure}

The suite of HST/WFC3 images obtained for NGC 6302 is presented in
Fig.~\ref{fig:NGC6302imageSuite}. Across this WFC3
image suite, as in the (WFPC2 and WFC3) images previously obtained by HST
\citep{Matsuura2005,Szyszka2009,Szyszka2011}, NGC 6302 displays its classical,
pinched-waist bipolar morphology, with its lobe-bisecting dark lane 
(the projection of its dusty torus against the sky) oriented
very nearly N--S \citep[see also][]{Rao2018}. The new WFC3 image suite 
clearly defines the wavelength-dependent morphologies of the dusty toroidal equatorial structure (central
dark lane) that bisects the polar lobes, as well as the fine structures (clumps, knots, and filaments) within the lobes.
On large ($\sim$60--90$''$; $\sim$0.3--0.45 pc) scales, the two large, ``major'' lobes are oriented more or 
less E--W, i.e., orthogonally to the dusty equatorial torus. However, it also evident that the lobes contain a complex system of point-symmetric sets of knot and filament structures spanning a broad ranges in both size scale and position angle (i.e., PAs from $\sim$20$^\circ$ to $\sim$150$^\circ$), and that the relative strengths of these lobe features vary sharply with wavelength and from emission line to emission line. 

Undoubtedly the most striking and unexpected aspect of the new WFC3
image suite is the bright, S-shaped 1.64 $\mu$m [Fe {\sc ii}] emission
that traces the southern interior of the east lobe rim 
and the northern interior of the west lobe rim, in point-symmetric
fashion (Fig.~\ref{fig:NGC6302imageOverlay1}). Fainter [Fe {\sc ii}] emission 
is also detected in the lobe interiors, especially to the northeast of the dark lane, 
as well as in the lobe ``clump zones'' (\S~\ref{sec:clumps}).  This constitutes the
first detection of 1.64 $\mu$m [Fe {\sc ii}] emission in NGC 6302; although the line is
evidently very bright (peak surface brightness $\sim$$3\times10^{-14}$
erg cm$^{-2}$ s$^{-1}$ ster$^{-1}$), previous near-IR spectroscopy of
NGC 6302 had not covered this (H band) spectral region
\citep{AshleyHyland1988,Casassus2000,Casassus2002}. The presence and
morphology of 1.64 $\mu$m [Fe {\sc ii}] emission is indicative of the
presence of off-axis shocks generated by fast ($\gtrsim100$ km s$^{-1}$),
collimated winds from the central star(s) of NGC 6302, as is discussed in
detail in later sections of this paper.

It is furthermore readily evident that the
surface brightnesses of the high-excitation species [Ne~{\sc v}] and [O~{\sc iii}] and the H recombination lines
(H$\beta$,  H$\alpha$, Pa$\alpha$) fall off more steeply with
increasing displacement from the pinched waist than those of
[N {\sc ii}] and [S {\sc ii}], with the latter, lower-excitation lines
conspicuously bright in the nebula's clump and knot structures and
remaining bright out to large displacements ($>$2$'$, i.e., $>$0.6 pc)
in the polar lobes. These gradients in nebular ionization and
excitation are illustrated in the form of RGB image overlays
in Fig.~\ref{fig:NGC6302imageOverlay2}, and are even more
apparent in the line ratio images presented in \S \ref{sec:lineRatios}. The top panel of
this Figure shows that [Ne {\sc v}] emission appears somewhat more
compact than [O {\sc iii}], with the caveat that the F343N [Ne {\sc v}] filter is
more susceptible than F502N  [O {\sc iii}] to contamination by
scattered starlight in the dusty lobe interiors. This color overlay also reveals
small-scale excitation gradients in the clump regions of the lobes (which are discussed in more
detail in \S \ref{sec:clumps}), illustrating in particular how the  [O
{\sc iii}] follows the clump exteriors, while Pa$\alpha$
emission effectively traces ionized regions that are highly
dust-extincted. The bottom frame of Fig.~\ref{fig:NGC6302imageOverlay2}
furthermore demonstrates the dramatically different extents of [Ne {\sc v}]
vs.\ [S {\sc ii}] emission and the strikingly different
morphologies of [S {\sc ii}] and [Fe {\sc ii}]. 

\section{Line ratio images}

\subsection{H recombination lines: mapping extinction}\label{sec:ext}

\begin{figure}[!ht]
\centering
\includegraphics[width=6in]{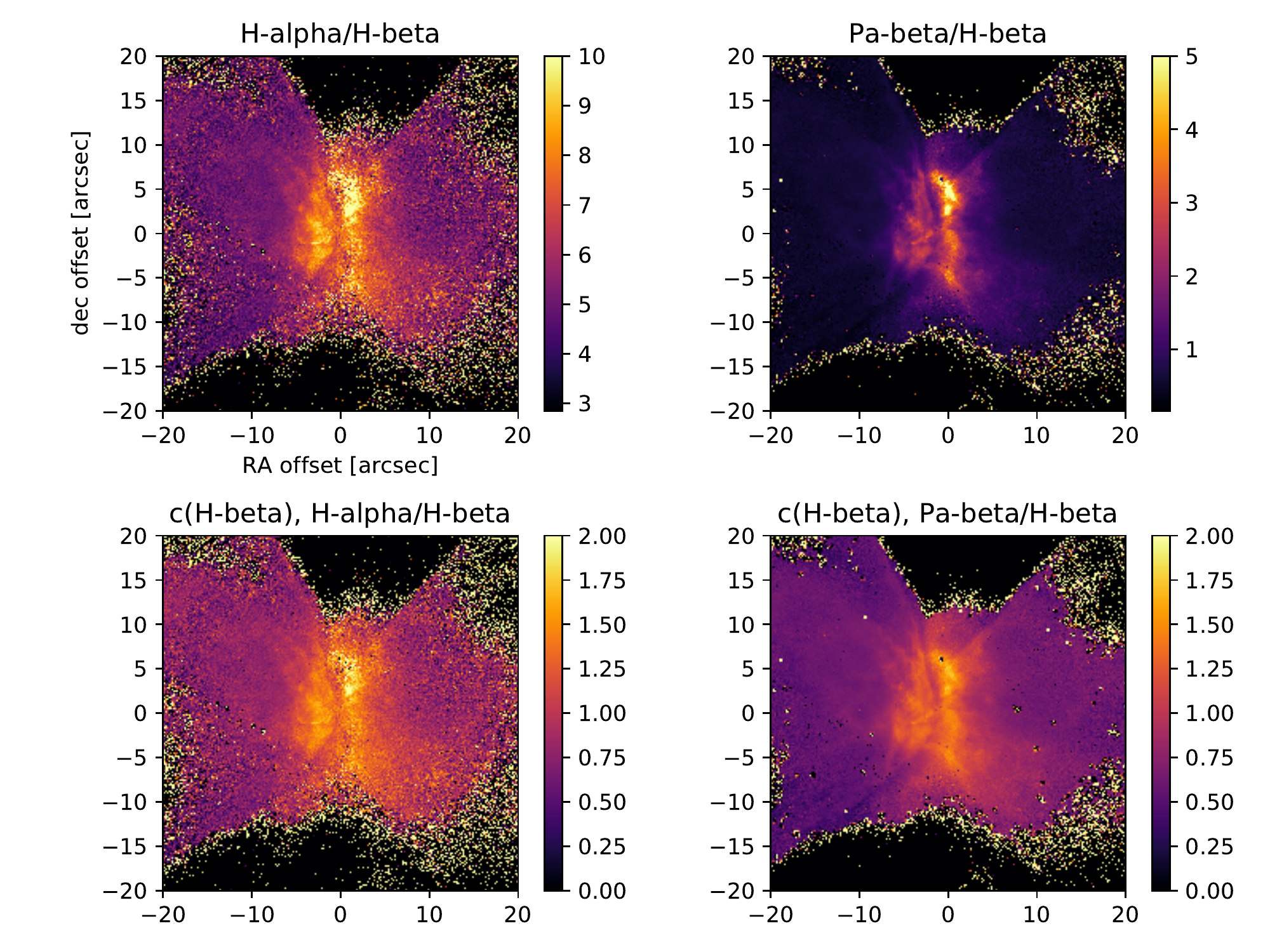}
\caption{
Top row: H$\alpha$/H$\beta$ (left) and Pa$\beta$/H$\beta$ (right) line
ratio images of NGC 6302. Bottom row:
maps of H {\sc i} line decrement extinction parameter ($c$) constructed from H$\alpha$/H$\beta$ (left) and
Pa$\beta$/H$\beta$ (right) line ratio images. Field of view is $40'' \times 40''$; N is up and E is to the left.
}
\label{fig:NGC6302extMaps}
%\vspace{-.25in}
\end{figure}

In Fig.~\ref{fig:NGC6302extMaps} (top panels), we present H recombination line
ratio images (H$\alpha$/H$\beta$ and Pa$\beta$/H$\beta$) of the
central $40''\times40''$ region of NGC 6302. In the regimes of
electron temperature and density appropriate to NGC 6302
\citep[i.e., $\sim$1.5$\times10^4$ K and $\sim$$10^4$ cm$^{-3}$, respectively;][]{Rauber2014}, the theoretical (Case B)
line ratios are H$\alpha$/H$\beta$ $\sim$2.85 and Pa$\beta$/H$\beta$
$\sim$0.162 \citep{Osterbrock2006}. It is readily apparent from
Fig.~\ref{fig:NGC6302extMaps}  that the observed ratios in NGC 6302
are everywhere larger than the theoretical ratios, by factors ranging from
$\sim$2 in the lobes to as large as $\sim$30, in the case of
Pa$\beta$/H$\beta$ in the central torus (dark lane) region. These large
deviations are caused by reddening that, in turn, is due to a
combination of foreground (ISM) and intranebular dust
extinction. Whereas ISM reddening 
\citep[estimated at $E(B-V) \approx 0.3$ toward NGC 6302;][]{Lallement2019}
likely dominates the (relatively smooth) extinction toward
the lobe regions, intranebular extinction clearly becomes the dominant
extinction source in the central torus region \citep[similar
conclusions were reached in previous studies of
NGC 6302; e.g.,][]{Matsuura2005,Rauber2014}. This contrast is
perhaps best seen in the Pa$\beta$/H$\beta$ image; the factor $\sim$3 wavelength range
spanned by this ratio makes it a sensitive probe of the highly
reddened central regions of NGC 6302.

Maps of the nebular extinction parameter $c$ obtained from the H$\alpha$/H$\beta$ and Pa$\beta$/H$\beta$ line
ratio images are presented in the bottom panels of
Fig.~\ref{fig:NGC6302extMaps}. These maps were generated from the
relation $c = [\log{(R_c/R_o)}]/[f(\lambda_2) - f(\lambda_1)]$  \citep[e.g.,][]{Groves2002}, where $R_c$
and $R_o$ are the theoretical and observed line ratios and
$f(\lambda_1), f(\lambda_2)$  are extinction values at the relevant
wavelengths as given by a reddening law. Here, we
adopted the ISM reddening law determined by \citet[][]{Cardelli1989}.

The detailed morphologies of the resulting $c$ maps in Fig.~\ref{fig:NGC6302extMaps} are fully consistent. However,
the values of $c$ in the map derived from Pa$\beta$/H$\beta$ are
systematically smaller than those derived from H$\alpha$/H$\beta$;
the map obtained from Pa$\beta$/H$\beta$ ranges from $c \sim 0.6$
(lobes) to a peak of $c \sim 1.7$ within the central waist
region, compared with the corresponding range of $c \sim$ 0.8--2.0 for
the map obtained from H$\alpha$/H$\beta$. This discrepancy may reflect
deviations from ``standard'' ISM dust extinction behavior as
encoded in the reddening law from \citet[][]{Cardelli1989}. 
% Furthermore, given the systematic nature of the discrepancy across the nebula, the
% discrepancy is more likely due to the properties (size
% distributions and/or scattering efficiencies) of the ISM dust grains along the
% line of sight toward NGC 6302, as opposed to the properties of the
% dust within the nebula. 

We note that the ranges of $c$ we derive from our HST/WFC3 H recombination line
ratio images (Fig.~\ref{fig:NGC6302extMaps}) are in reasonable agreement with previous
determinations for $c$ in NGC 6302 as obtained from long-slit
spectroscopy, aperture spectroscopy, and/or
aperture photometry at wavelengths ranging from the visible to the
radio, i.e.,  $c \sim 0.8$ in the lobes and $c \sim 1.2$ to $\sim$2.4
in the central torus region \citep[][and references
therein]{Groves2002,Rauber2014}. At the same time it is readily apparent, given
the very large range in $c$ observed in our extinction maps, that previous
measurements would be tremendously sensitive to the specifics of aperture placement,
wavelength span, and assumed reddening law. This of course has profound implications for
determinations of line ratios and (hence) ionic abundances,
temperatures, and densities within the nebula  \citep[see
also][]{Rauber2014}.

\begin{figure}[!ht]
\centering
\includegraphics[width=4in]{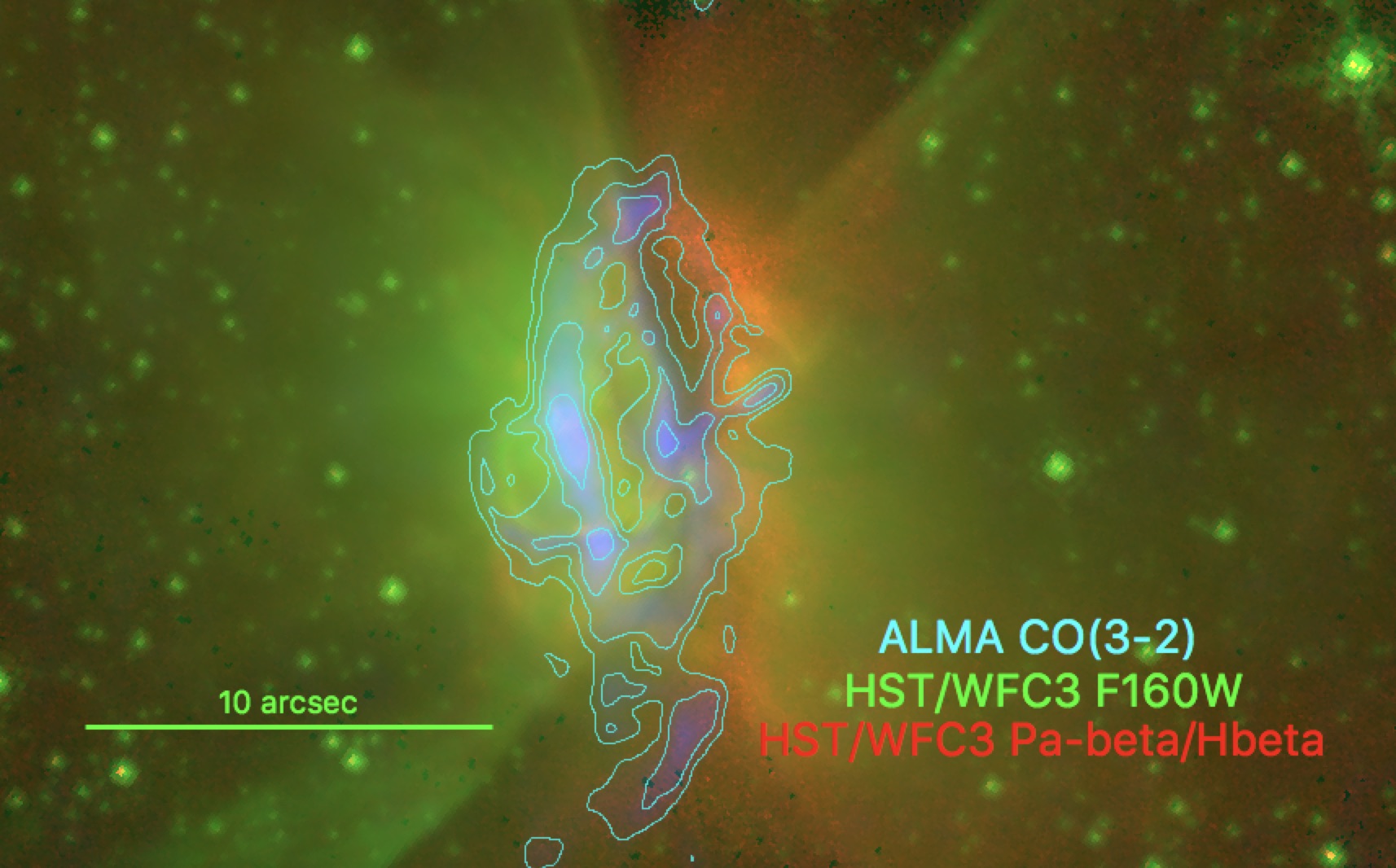}
\caption{
Color overlay of ALMA map of 345 GHz
CO $J=3\rightarrow2$  emission (blue and contours) from \citet{SantanderGarcia2017} on HST/WFC3 Pa$\beta$/H$\beta$ line ratio
and F160W (``H band'') images (red and green, respectively). The Pa$\beta$/H$\beta$ ratio image is displayed on a square-root intensity scale, 
with a range of 0.16--6.8. The CO contour
levels are 0.025, 0.4, 0.8, 1.5, and 2.5 Jy km s$^{-1}$
beam$^{-1}$, and the beamsize is 0.36$''$. Field of view is $35''\times 20''$; N is up and E is to the left.
}
\label{fig:NGC6302extCO}
%\vspace{-.25in}
\end{figure}

In Fig.~\ref{fig:NGC6302extCO} we present a color overlay of the ALMA
integrated intensity map of 345 GHz CO $J=3\rightarrow2$ emission from
\citet{SantanderGarcia2017} on our HST/WFC3 Pa$\beta$/H$\beta$ line
ratio and F160W (``H band'') images, zoomed in on the core region of the nebula. 
As described in detail in \citet{SantanderGarcia2017}, the main bright loop of CO traces the
molecular gas within the central, expanding, dusty torus that obscures
the CSPN of NGC 6302. 
While the overall correspondence between
the surface brightness of CO emission and the distribution of dust
extinction (as mapped by Pa$\beta$/H$\beta$) is apparent in
Fig.~\ref{fig:NGC6302extCO}, the CO and dust extinction
morphologies clearly differ in detail. In particular, the western rim
of the CO loop, which is blueshifted \citep{SantanderGarcia2017},
precisely traces the sharp, narrow minimum in Pa$\beta$/H$\beta$ that bisects this
line ratio image. This spatial correspondence between blueshifted CO
and the Pa$\beta$/H$\beta$ ``dark lane'' indicates that the
forward-facing portion of the expanding CO loop traces a large
intervening column of dust --- a dust column large enough that
too few Pa$\beta$ line photons escape for the
Pa$\beta$/H$\beta$ ratio to serve as a probe of the dust extinction. The fact that the Pa$\beta$/H$\beta$ ratio rises to
its maximum value ($\sim$2.0) at a displacement $\sim$2$''$ west of
this sharp minimum, before declining \citep[a result generally consistent with
that from long-slit optical spectral mapping;][]{Rauber2014}, then reveals
the radial extent of the dusty torus as seen in projection toward the west
lobe,  i.e., $\sim$2000 au. In contrast, the bright eastern rim of the CO loop, which is
redshifted \citep{SantanderGarcia2017},
has no clear counterpart in the Pa$\beta$/H$\beta$ line ratio image. This portion of the
central molecular torus is projected behind the east lobe, such that its dust
counterpart cannot be traced via extinction mapping.

\subsection{High- and low-excitation forbidden lines: ionization
  gradients} \label{sec:lineRatios}

\begin{figure}[!ht]
\centering
\includegraphics[width=6in]{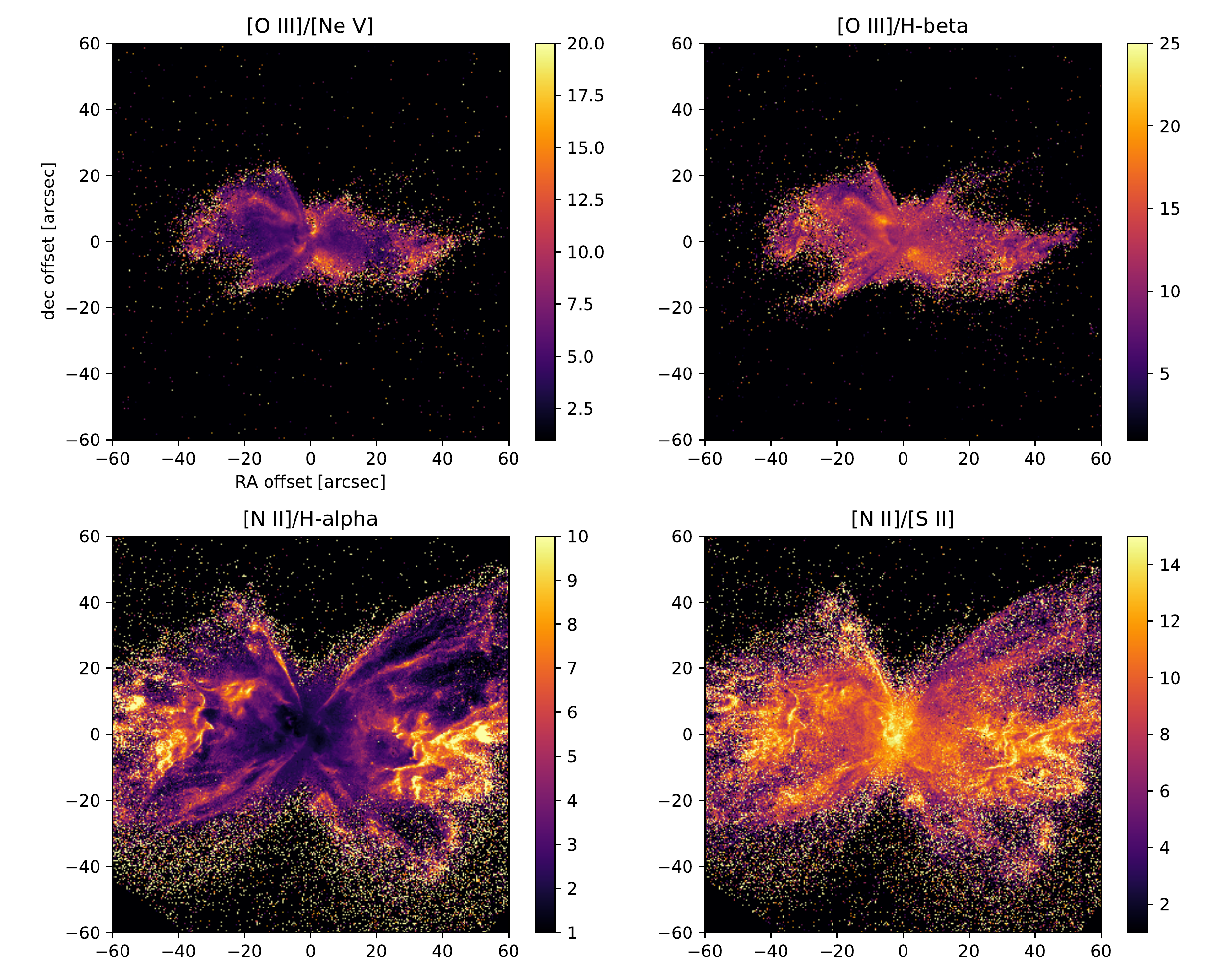}
\includegraphics[width=4.5in]{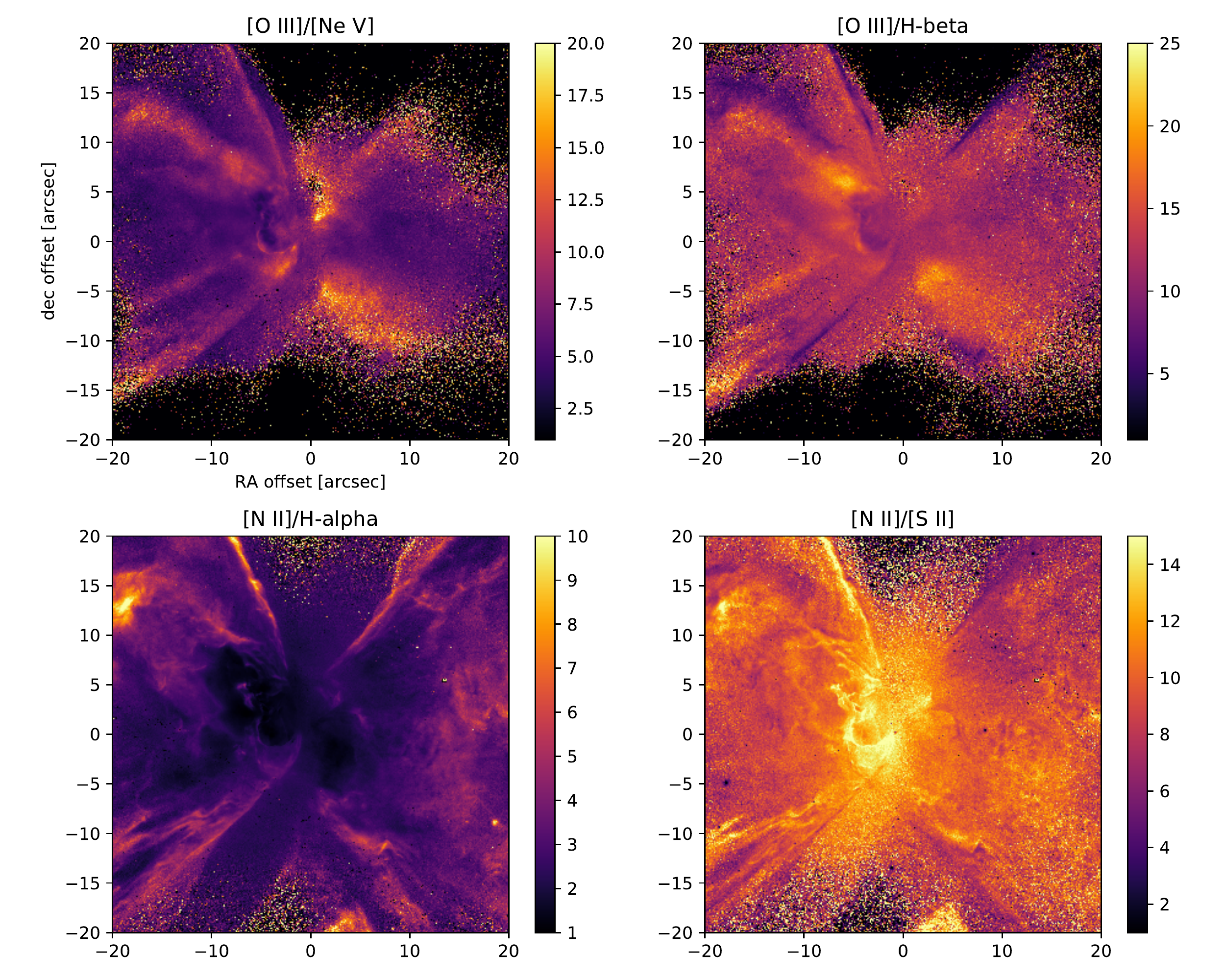}
\caption{
Various line ratio images of NGC 6302. In each set of four panels, [O {\sc iii}]/[Ne {\sc v}]  (F502N/F343N; left) and [O {\sc iii}]/H$\beta$ (F502N/F487N; right) are displayed in the top row, and [N {\sc ii}]/H$\alpha$ (F658N/F656N; left) and [N {\sc ii}]/[S {\sc ii}] (F658N/F673N; right) are displayed in the bottom row. Fields of view are $120'' \times 120''$ (top panels) and $40'' \times 40''$ (bottom panels); N is up and E is to the left.
}
\label{fig:NGC6302ratioMaps}
%\vspace{-.25in}
\end{figure}

\begin{figure}[!ht]
\centering
\includegraphics[width=6in]{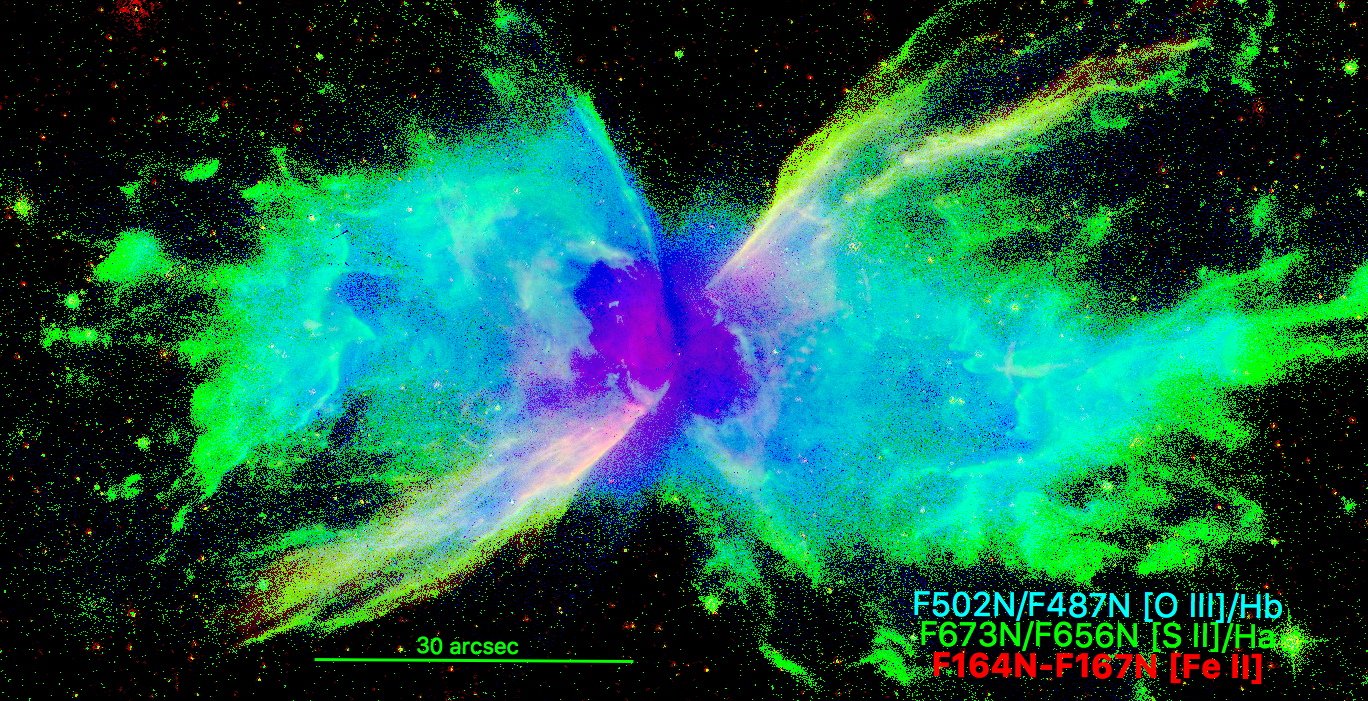}
\includegraphics[width=6in]{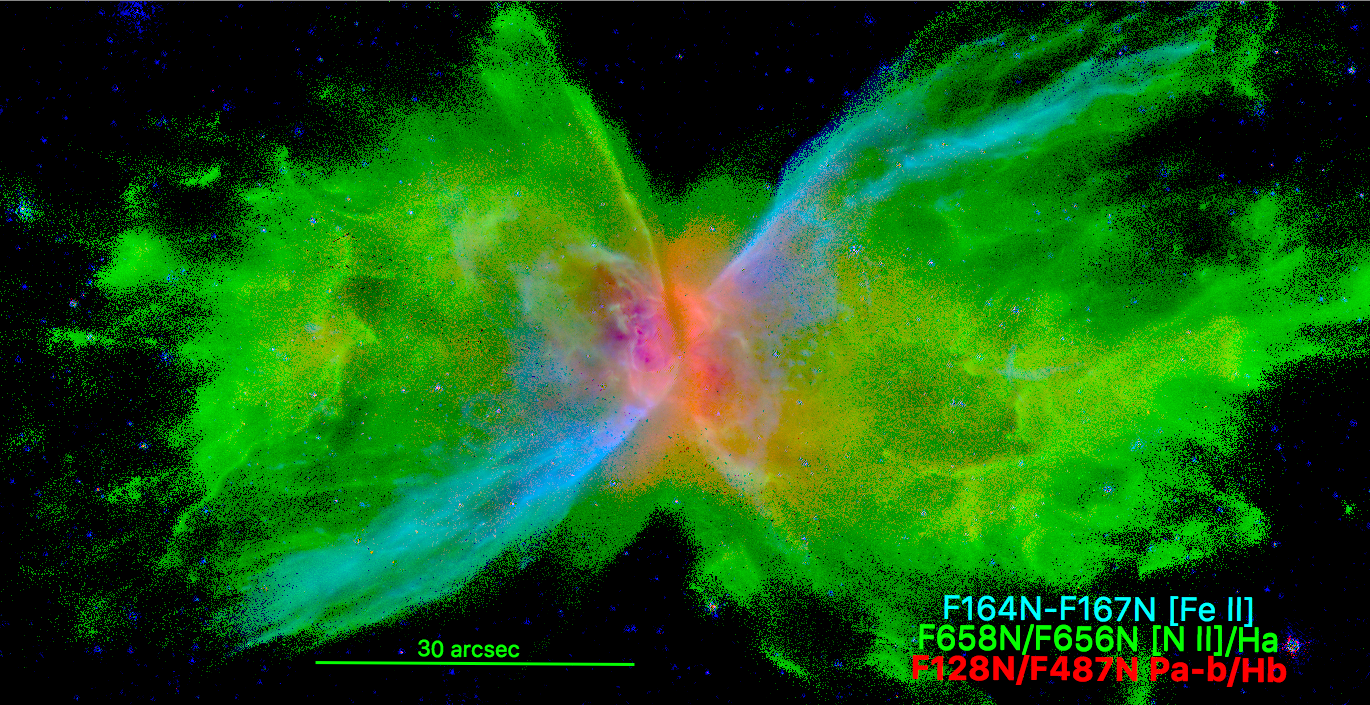}
\caption{Color overlays of WFC3 line
ratio and 1.64 $\mu$m [Fe {\sc ii}] images of NGC 6302, rendered in log scale. Top: [O {\sc iii}]/H$\beta$ (F502N/F487N; range 1.0--30), 
[S {\sc ii}]/H$\alpha$ (F673N/F656N; range 0.2--1.0), and [Fe {\sc ii}]
(F164N$-$F167N difference image) coded blue, green, and red, respectively. Bottom:  [Fe {\sc ii}], [N {\sc ii}]/H$\alpha$
(F658N/F656N; range 1.0--30), and Pa$\beta$/H$\beta$ (F128N-F130N divided
by F487N; range 0.5--5.0) coded blue, green, and red, respectively. Field of view is $120'' \times 70''$; N is up and E is to the left. 
%The white dashed lines superimposed in each frame indicate the positions and orientations of the apertures used to extract the spatial line ratio profiles displayed in Figure~{\bf XXX}. }
}
\label{fig:NGC6302ratioOverlay}
%\vspace{-.25in}
\end{figure}

\begin{figure}[!ht]
\centering
\includegraphics[width=2.25in]{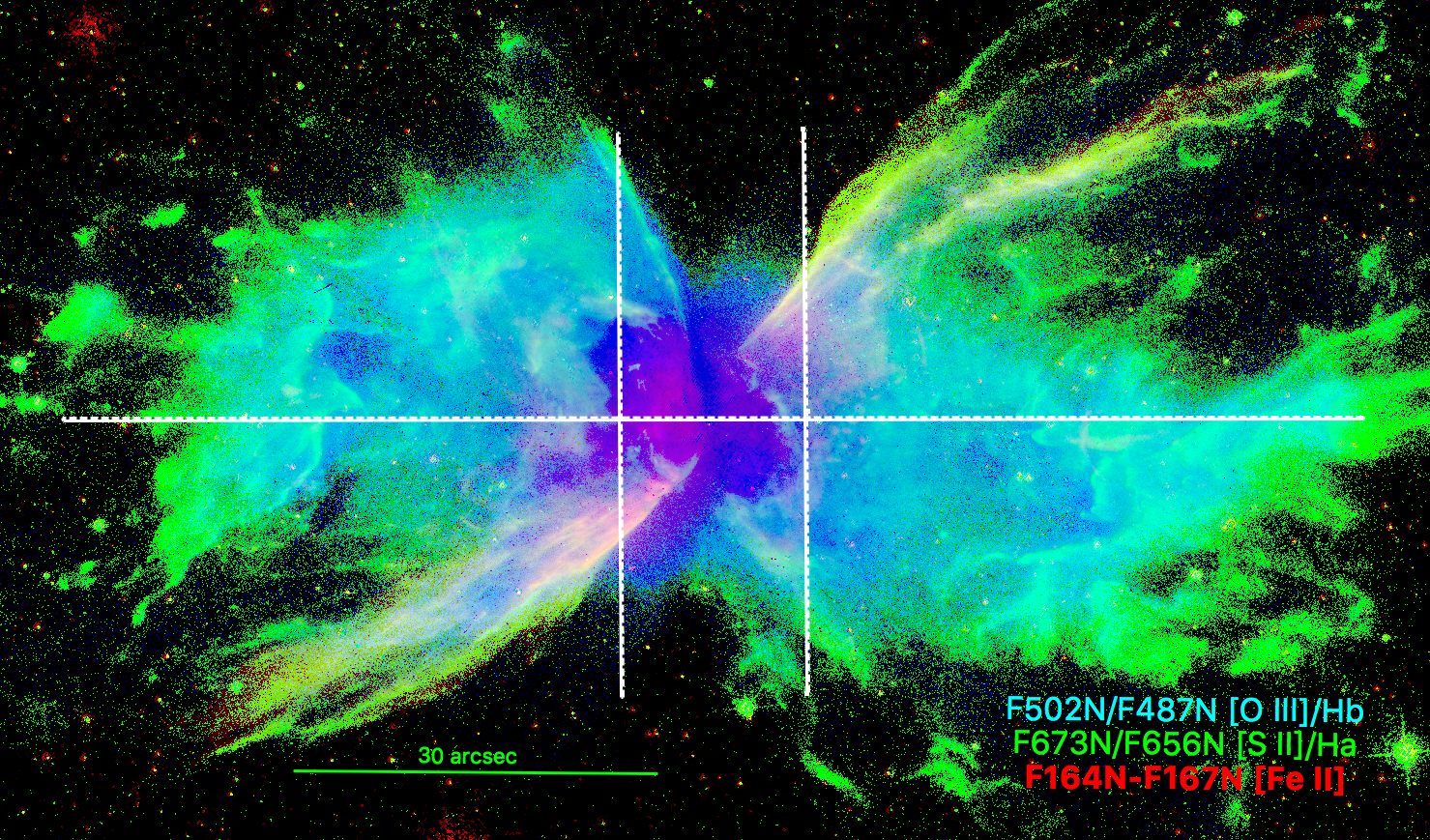}
\includegraphics[width=2.25in]{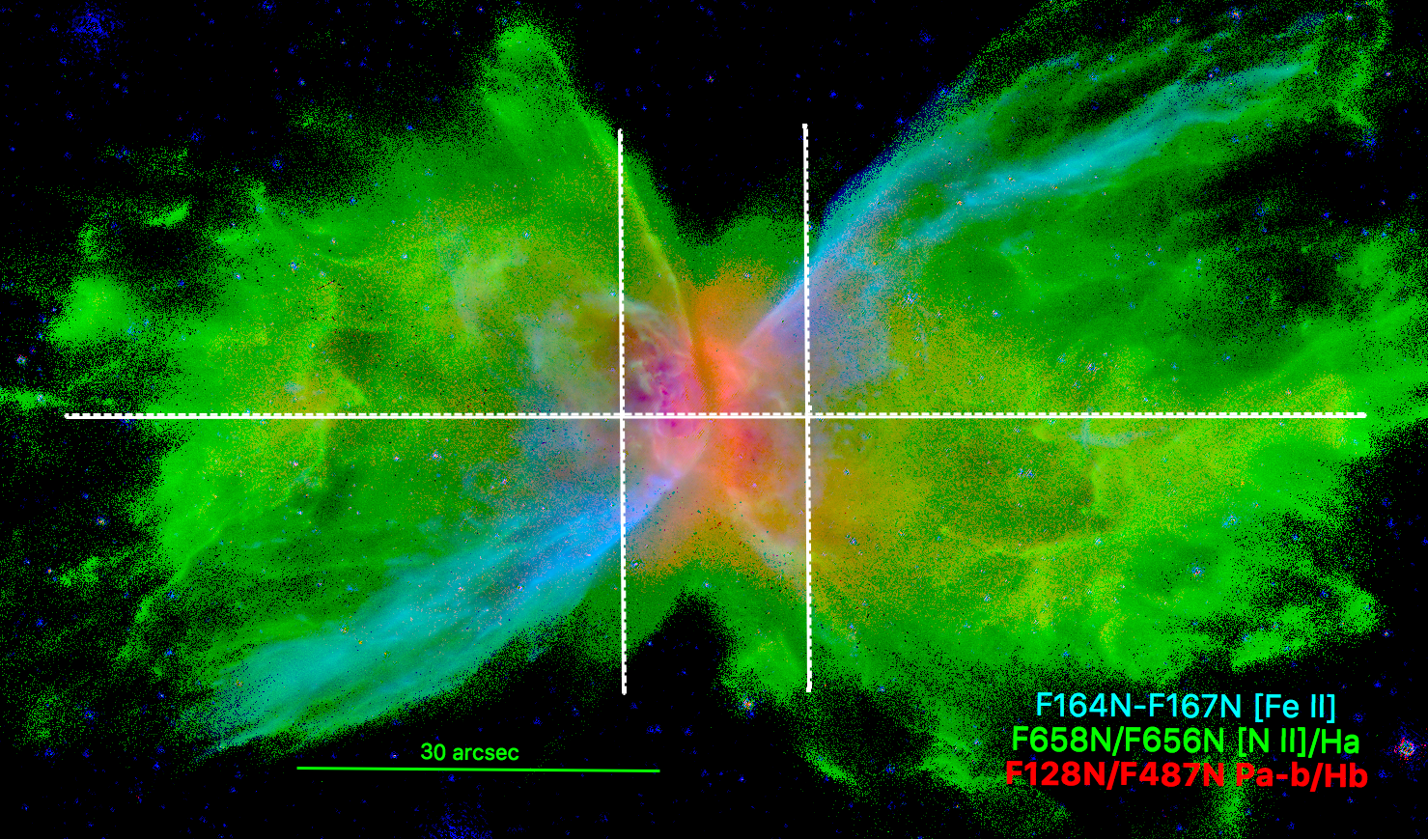}
\includegraphics[width=5in]{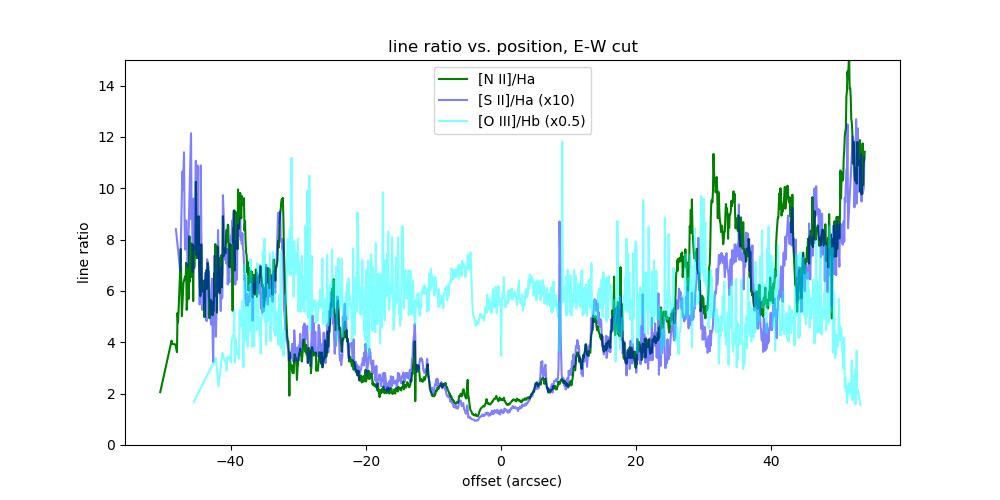}
\includegraphics[width=5in]{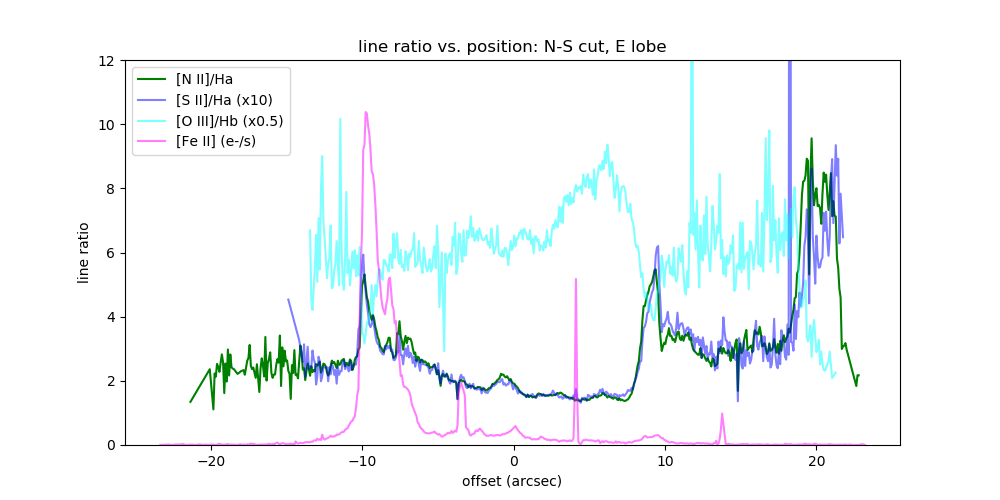}
\includegraphics[width=5in]{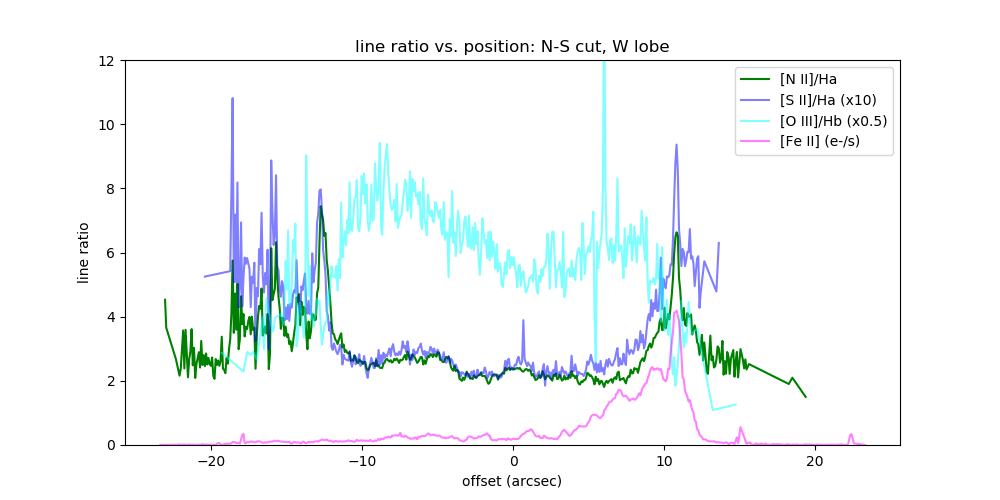}
\caption{Top panels: Color overlays of WFC3 line
ratio and 1.64 $\mu$m [Fe {\sc ii}] images of NGC 6302, from Fig.~\ref{fig:NGC6302ratioOverlay}, overlaid with white lines indicating the positions (lengths and orientations) of the (2$''$ wide) apertures used to extract the spatial line ratio profiles displayed in the next three panels. Middle upper panel: E-W cut through the inferred position of the central star (see \S~\ref{sec:CSPN}), with positive (negative) indicating offset W (E) of this position. Middle lower panel: N-S cut through the E lobe. Bottom panel: N-S cut through the W lobe.
}
\label{fig:NGC6302ratioCuts}
%\vspace{-.25in}
\end{figure}

In Fig.~\ref{fig:NGC6302ratioMaps}, we present a series of line ratio
images obtained from high-excitation forbidden lines ([Ne~{\sc v}],
[O~{\sc iii}]) and low-excitation forbidden lines ([S~{\sc ii}], [N~{\sc ii}])
with respect to each other and to H recombination lines. Of these line ratio images, only [O~{\sc iii}]/[Ne~{\sc v}] should be
sensitive to extinction. Given the results above (\S~\ref{sec:ext}), however, this ratio should only be affected by extinction
within the central torus region, where contamination of the F343N [Ne~{\sc v}] image by continuum and O {\sc iii} Bowen fluorescence line
emission is also likely most significant \citep[see][their Fig.~2]{Groves2002}. 
We note that the (relatively broad) F343N is the only UVIS filter that should be affected by such contamination due to continuum or line emission.
%The [O~{\sc iii}]/H$\beta$, [S~{\sc ii}]/H$\alpha$, and [N~{\sc ii}]/H$\alpha$ 

The [O~{\sc iii}]/H$\beta$, [S~{\sc ii}]/H$\alpha$, and [N~{\sc ii}]/H$\alpha$ line ratio images, along with Pa$\beta$/H$\beta$ and [Fe~{\sc ii}] images 
(as nebular extinction and shock tracers, respectively; see \S~\ref{sec:ext} and below), 
are also presented in the form of color overlays in Fig.~\ref{fig:NGC6302ratioOverlay}. 
In Fig.~\ref{fig:NGC6302ratioCuts}, we display profiles of the values of the [O~{\sc iii}]/H$\beta$, [S~{\sc ii}]/H$\alpha$, and [N~{\sc ii}]/H$\alpha$ line ratios 
as measured along (three) selected linear cuts through the line ratio images 
--- one cut along the E-W direction through the inferred position of the central star (see \S~\ref{sec:CSPN}), and two N-S cuts through the inner lobe regions ---
with the positions and lengths of these cuts indicated in the color overlays at the top of the Figure.

Broadly speaking, the ionization structure observed in these HST/WFC3 line ratio
maps --- wherein the higher ionization state species  (Ne$^{4+}$,
O$^{2+}$) are brightest in the lobe interiors, and the lower
ionization species (singly ionized S and N) become most prominent in
the outer reaches of the nebula, relative to H$\beta$ and H$\alpha$
(respectively) --- is consistent with the predictions
of (e.g., CLOUDY) models of planetary nebulae with simple (spherical)
geometries that are photoionized by hot central stars
\citep[e.g.,][]{AlexanderBalick1997}. This ionization structure is readily apparent in Fig.~\ref{fig:NGC6302ratioOverlay}, upper panel, 
and in the E-W cut through the polar lobes displayed in Fig.~\ref{fig:NGC6302ratioCuts}. However --- even accounting for the
likely additional contribution of shocks \citep{Lago2019} --- the
line ratio images clearly deviate from such simple prescriptions in
fundamental respects, which we highlight here.

The most striking aspect of the line ratio images involving the high-excitation
lines is the series of radially directed, azimuthally alternating high/low ratio
features, extending away from the central waist
region. These features are most noticeable within the east lobe, where 
they more or less coincide with the boundaries of the distinct
azimuthal regions (wedges) discussed in detail in Sec.~\ref{sec:wedges}. The  [O~{\sc iii}]/H$\beta$ ratio lies in the
range $\sim$10--15 across much of the lobe regions (Fig.~\ref{fig:NGC6302ratioCuts}), consistent with
previous spectroscopic mapping within a $\sim$$200'' \times 10''$
strip oriented along the polar axis of NGC 6302 by \citet{Rauber2014}, and as
expected for high-excitation nebulae \citep[e.g.,][]{AlexanderBalick1997}. However, the [O {\sc iii}]/H$\beta$
ratio is observed to climb to $\sim$15--20 in narrow strips oriented along PAs $\sim$60$^\circ$ and
$\sim$120$^\circ$, and drops to distinct minima of [O {\sc
  iii}]/H$\beta$ $\sim$6--8 along PAs $\sim$20$^\circ$,
$\sim$45$^\circ$, and $\sim$150$^\circ$.  
These strips of elevated and depressed [O~{\sc iii}]/H$\beta$ ratio 
can be seen as peaks and valleys in the N-S cut through the east lobe displayed in Fig.~\ref{fig:NGC6302ratioCuts}. 
Radially aligned  (azimuthally alternating) features of locally high and low [O
{\sc  iii}]/H$\beta$ ratio are also seen in the west lobe, but are
generally less distinct, with the exception of narrow radial strips of
low [O~{\sc iii}]/H$\beta$ ($\sim$5) along the lobe perimeters (PAs $\sim$200$^\circ$ and
$\sim$315$^\circ$). The N-S profiles displayed in Fig.~\ref{fig:NGC6302ratioCuts} (lower panels) 
furthermore demonstrate that the minima in [O~{\sc iii}]/H$\beta$ along the southern and northern perimeters of 
the east and west lobes, respectively, correspond to sharp peaks in [Fe~{\sc ii}] surface brightness.

The [O {\sc iii}]/[Ne {\sc v}] ratio image morphology closely follows
that of [O {\sc iii}]/H$\beta$, wherein [O {\sc iii}]/[Ne {\sc v}] is
somewhat elevated within the radial features that display larger [O {\sc iii}]/H$\beta$. Since
the ionization potential of Ne$^{3+}$ is much larger than that of
O$^{+}$ (97 eV vs.\ 35 eV), this suggests that the azimuthal regions
characterized by low [O {\sc iii}]/H$\beta$ and [O {\sc iii}]/[Ne {\sc
  v}] are in fact the highest-excitation regions of the nebula; i.e.,
in these regions, where [Ne {\sc v}] is bright but [O {\sc iii}] is
relatively weak, O$^{2+}$ is suppressed in favor of higher ionization 
states of O. The implied large azimuthal ionization gradients are impossible to reconcile with 
standard (Str\"{o}mgren-sphere-based) ionization theory, and point
instead to shadowing effects. That is, the lobe regions with
supressed [O {\sc iii}]/H$\beta$ and [O {\sc iii}]/[Ne {\sc 
  v}] are likely directly exposed to EUV and soft X-ray radiation 
from NGC 6302's exceedingly hot central star; whereas the radial zones
of elevated [O {\sc iii}]/H$\beta$ and [O {\sc iii}]/[Ne {\sc 
  v}] mark regions where radiation from the central star
is at least partially attenuated, perhaps by intervening dust. 

The structures observed in the line ratio images constructed from [N~{\sc ii}]
and [S~{\sc ii}] differ fundamentally from those of the
high-excitation forbidden lines. In particular, as is evident from the profiles displayed in Fig.~\ref{fig:NGC6302ratioCuts}, 
these low-excitation lines are particularly bright (with respect to H$\alpha$) in the clump and ``elephant trunk'' structures
within the lobe interiors at offsets of $\sim$30$''$--60$''$ from the
core region. These regions are discussed in detail in \S~\ref{sec:clumps}. In the core region, we find [N {\sc ii}]/H$\alpha$ $<3$, 
whereas within the clump regions of the lobe
interiors, the [N~{\sc ii}]/H$\alpha$ ratios climb to $\sim$10. The lobe perimeters, which are marked by very sharp
gradients in [O~{\sc iii}]/H$\beta$ ratio, appear as narrow
regions of enhanced [N~{\sc ii}]/H$\alpha$ and [S~{\sc ii}]/H$\alpha$. 
The profiles in Fig.~\ref{fig:NGC6302ratioCuts} also demonstrate that the [N~{\sc ii}]/[S~{\sc ii}] line ratio is $\sim$10 throughout much of the nebula, 
with the exception of the core region and the west lobe clumps and perimeter regions. The core region is both deficient in low-excitation 
forbidden line emission and relatively high in [N~{\sc ii}]/[S~{\sc ii}] line
ratio ([N~{\sc ii}]/[S~{\sc ii}] $\sim$ 15). The clumps and perimeter regions in the west lobe also display [N {\sc ii}]/[S {\sc ii}] ratios of $\sim$15--20.
As is the case for [O~{\sc iii}]/H$\beta$, these [N {\sc ii}]/H$\alpha$ and [S~{\sc ii}]/H$\alpha$ (hence [N {\sc ii}]/[S {\sc ii}]) line ratio
mapping results are quite consistent with those presented in
\citet{Rauber2014}, but obviously our images expand the line ratio
coverage to the entire surface area of both lobes, and at subarcsecond
resolution.

As discussed in detail in \citet{Lago2019}, the unusually large  [N {\sc
  ii}]/H$\alpha$ ratios of $\sim$10 measured in the lobe clump regions are
indicative of the influence of shocks on the observed ionization 
structure of NGC 6302. Indeed, there is ample
evidence for the presence of relatively strong ($\gtrsim$100 km s$^{-1}$) shocks in NGC
6302, both on the basis of large- and small-scale nebular morphology
and in the form of the bright 1.64 $\mu$m [Fe {\sc ii}] emission discovered in our
HST/WFC3 imaging (both aspects are discussed further in
Sec.~\ref{sec:wedges}). 
Such fast shocks might explain the large observed  [N {\sc
  ii}]/H$\alpha$ in the clumps as well as the elevated  [N {\sc
  ii}]/[S {\sc ii}] ratios in both the clump and core regions,
according to models describing Herbig-Haro objects associated with
collimated outflows from pre-main sequence stars
\citep[][]{Hartigan1994} --- assuming such models might be applicable to the extreme UV irradiation environment that characterizes NGC 6302.
However, the widely distributed lobe clump regions, where the
most highly elevated  [N~{\sc ii}]/H$\alpha$ ratios are observed, do not spatially coincide with
the (narrow) zones of bright 1.64 $\mu$m [Fe {\sc ii}] emission (Fig.~\ref{fig:NGC6302ratioOverlay},
lower panel). Furthermore, models specifically formulated to describe shocked ``cloudlets'' in planetary nebulae do not achieve such
large  [N {\sc  ii}]/H$\alpha$ ratios, although these models can reproduce the  [N {\sc ii}]/[S {\sc ii}] ratios we measure in the  clump and core regions
of NGC 6302 \citep{Raga2008}. 

\section{The azimuthal lobe structures of NGC 6302}  \label{sec:wedges}

\begin{figure}[!ht]
\centering
\includegraphics[width=7in]{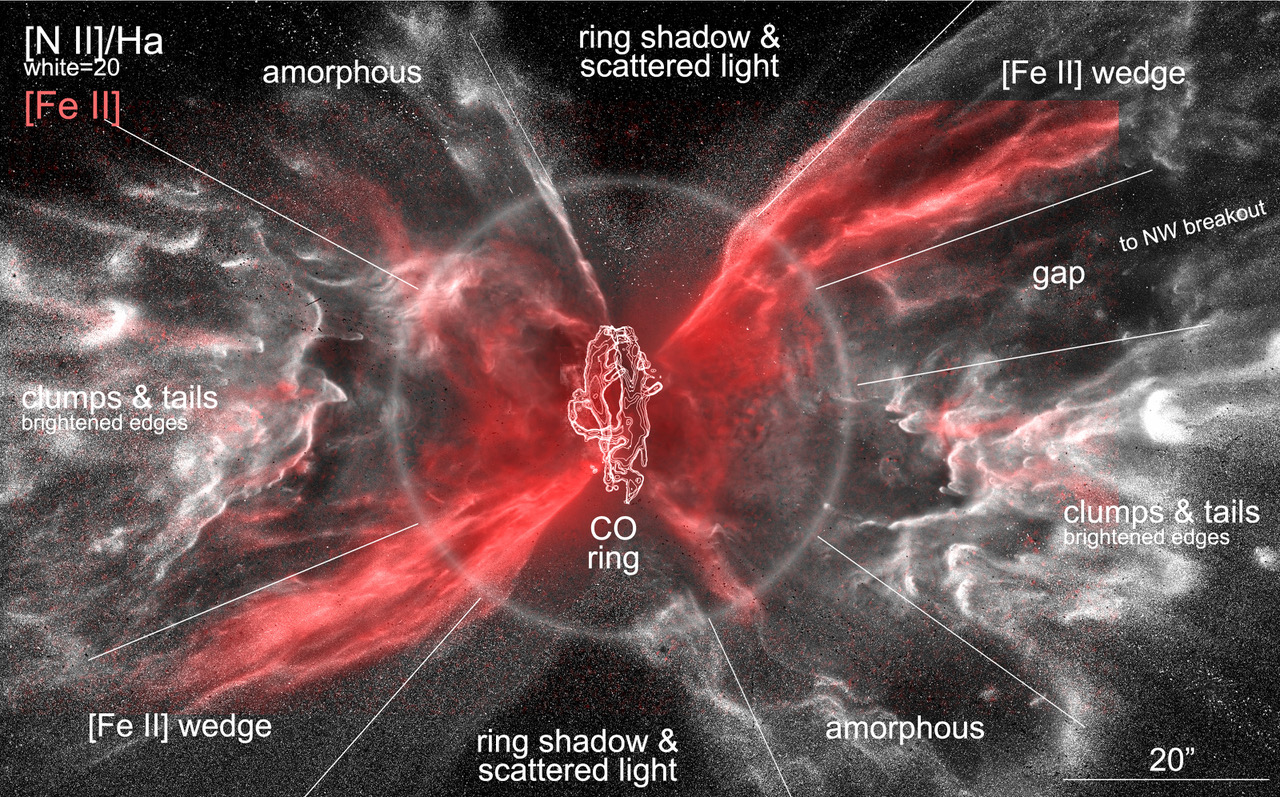}
\caption{
Overlay of [O {\sc iii}]/H$\beta$ ratio image (greyscale) and [Fe {\sc ii}] image
(red), annotated to indicate the various azimuthally organized zones (wedges)
of NGC 6302 that are described in the text (\S~\ref{sec:wedges}). Contours of the ALMA map
of CO emission from the central torus region \citep{SantanderGarcia2017} are overlaid in white. The $\sim$$20''$ radius
white circle delineates the approximate extent of the nebular core region.
}
\label{fig:NGC6302zones}
%\vspace{-.25in}
\end{figure}

%Morphologically, NGC 6302 is among the most complex of bipolar PNe.
%The new HST/WFC3 image suite allows us to begin to dissect its intricate structure. 
As illustrated in Fig.~\ref{fig:NGC6302zones}, the regions beyond the central nebular core region 
of NGC6302 (whose extent is roughly delineated by the white circle in the Figure) can be divided into distinct azimuthal wedges that extend radially outward from the core.
As noted in \S~\ref{sec:lineRatios}, the boundaries
of these wedge-shaped regions (indicated in Fig.~\ref{fig:NGC6302zones} as thin white lines extending radially away from the core region) are marked by strong azimuthal
gradients in nebular ionization structure. Such a clear azimuthal segmentation of spatial structure and
ionization in the lobe interiors is highly unusual
in bipolar PNe, apart from certain multipolar PNe.   

The general structure of the azimuthal lobe domains is as follows. Nearest the E-W nebular symmetry axis lie oppositely positioned
wedge regions containing highly conspicuous clumps with radially-directed outward-pointing
tails.  Moving clockwise, we find opposing wedges,
marked ``amorphous'' in Fig.~\ref{fig:NGC6302zones}, that display a paucity of structure relative to
the adjacent E-W oriented clump zones.  The next
wedges, marked ``ring shadow \& scattered light,'' are dark. These wedges mark
azimuthal angles where the central dusty, equatorial torus evidently obscures the central star from the
standpoint of any nebular material that may be present beyond a few arcsec from
the central waist. This dark shadow zone is bordered by (what appears to be) faint dust-scattered scattered
nebular emission.  Continuing clockwise, the wedges containing the
plumes of [Fe~{\sc ii}] are found next. An additional wedge at PA
$\sim$$-60^\circ$ appears as a gap, and marks the inner regions of a
pc-scale ``breakout'' lobe to the west, i.e., a region of the W lobe
in which both the proper motions and the Doppler speed of lobe
widening are extreme \citep[e.g.,][]{Meaburn2005,Meaburn2008}. (The breakout lobe will be explored in a second
paper on nebular dynamics.) 

\subsection{The ``clump'' wedges} \label{sec:clumps}

\begin{figure}[!ht]
\centering
\includegraphics[width=7in]{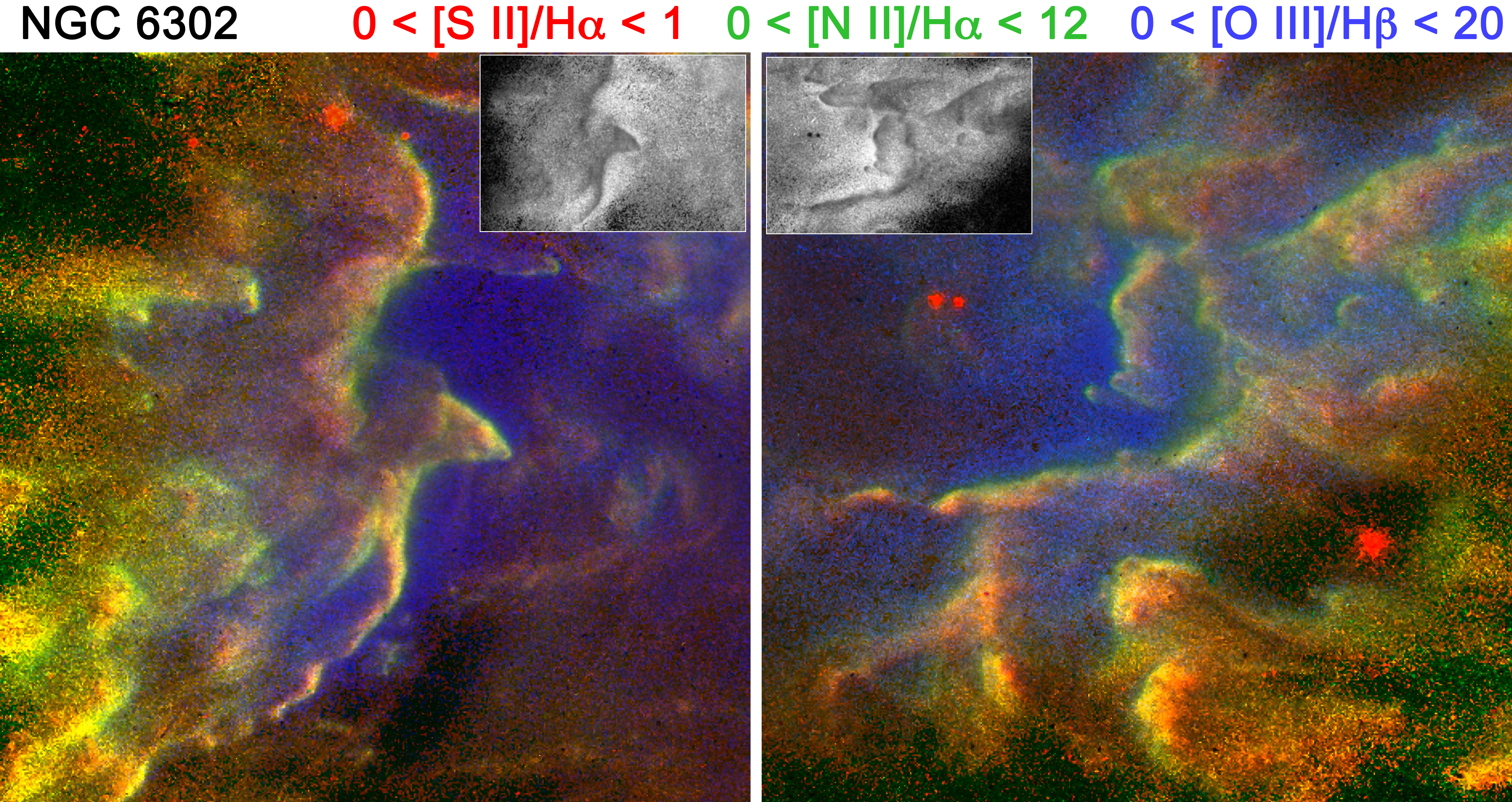}
\caption{
Color overlays of WFC3 line ratio images within $25'' \times 25''$ regions of NGC 6302  centered on the east (left) and west (right) clump zones. In the two panels, [O {\sc iii}]/H$\beta$ (F502N/F487N), [N
{\sc ii}]/H$\alpha$ (F658N/F656N), and [S
{\sc ii}]/H$\alpha$ (F673N/F656N) are coded blue, green, and red, respectively. The insets show greyscale representations of the central  $20'' \times 15''$ regions of the [O {\sc iii}]/H$\beta$ ratio image.
}
\label{fig:NGC6302clumps}
%\vspace{-.25in}
\end{figure}

Figure~\ref{fig:NGC6302clumps} illustrates the ionization structure and
selected line ratios for the E-W oriented ``clump'' wedges of
the nebula, via blowups of the brightest clump regions, exploiting line ratios that are
relatively insensitive to extinction. The line ratios 
are sensitive to both the degree of ionization and the local
excitation temperature.  The lobe clumps stand out in [N~{\sc ii}]/H$\alpha$ and (to a lesser extent) 
[S~{\sc ii}]/H$\alpha$ (yellow-green regions in color panels of Fig.~\ref{fig:NGC6302zones}) and are notably 
darkened in  [O {\sc~iii}]/H$\beta$ (Fig.~\ref{fig:NGC6302zones}, see greyscale insets).
The ratios of the optical forbidden lines to
neighboring Balmer lines (at adjacent wavelengths) are all enhanced on
the star-facing edges of the clumps, much like the clumps at the
perimeters of most photoionized H {\sc ii} regions and some wind-shocked
clumps seen in H-H objects \citep{Hartigan1994}. 

The ``leeward''
sides of the clumps (i.e., the sides facing away from the CSPN)
uniformly display tails, with widths similar to the clump diameters,
pointing away from the central star. 
These structures resemble (at least
superficially)  the dense, tail-bearing globules seen in dusty,
molecule-rich PNe like NGC 6720 \cite[the Ring;][]{Speck2003} and NGC 7293 \cite[the Helix;][]{Odell1996}. As in those PNe, the clump tails in NGC 6302 suggest that
radially directed CSPN winds with significant ram pressure are
sweeping clump material outwards. That is, the tails resemble wakes
that trail the clumps at their heads, indicating that they are immersed and entrained in a 
wind whose speed mildly exceeds the speed of the shock that strikes
the heads.  The fact that some of these head+tail features 
are traced by filamentary 1.64 $\mu$m [Fe~{\sc ii}] emission (Fig.~\ref{fig:NGC6302imageOverlay1}) 
supports this interpretation and,
furthermore, suggests that the velocities of the entraining 
winds, relative to the clump/tail structures, well exceed $\sim$50 km s$^{-1}$ (see \S \ref{sec:FeWedges}).
%Accordingly, we cannot presume that the clumps are entirely
%photoionized and heated. 

Four of our WFC3 images were obtained in filters that were also used in previous
(epoch 2009) HST/WFC3 imaging of NGC 6302  (specifically, F502N,
F656N, F658N, and F673N). The {\tt AstroDrizzle} image registration procedure described in \S~2, 
which exploits field stars common to image pairs, was used to place these
WFC3 images into the astrometric coordinate frame of the images in the 
2019--2020 WFC3 image suite that were obtained through the corresponding filters. 
Detailed analysis of the nebular kinematics
revealed in difference and ratio images contructed from these
multi-epoch images is deferred to a subsequent
paper (Balick et al., in preparation). Here we merely note that (a) the lobe clump/tail
structures generally display the clearest signs of
proper motion of any nebular structures, due to their well-defined head/tail
structures; and (b) our difference-image-based analysis of the ballistic expansion of these structures, 
analogous to that described in \citet{Schoenberner2018}, 
indicates that their characteristic dynamical (ejection) timescale 
is $\sim$2000 yr. The latter result is consistent with previous multi-epoch imaging
studies of the nebula \citep{Meaburn2008,Szyszka2011}. However, some inner lobe structures also
clearly display proper motions that are of similar magnitude to those
of the outer lobe structures, suggestive of more recent mass ejections (see \S \ref{sec:CSPN}).

Beyond the clumps’ bright edges 
and along the outer edges of their tails, the [O {\sc iii}]/H$\beta$ ratio drops
precipitously (see Fig.~\ref{fig:NGC6302ratioCuts}, E--W profile, and 
Fig.~\ref{fig:NGC6302clumps}, greyscale insets).
In contrast, the [N~{\sc ii}]/H$\alpha$ and [S~{\sc ii}]/H$\alpha$ ratios both significantly
increase within the clumps (again see Fig.~\ref{fig:NGC6302ratioCuts}, E--W profile, and Fig.~\ref{fig:NGC6302clumps}, color panels).
On the one hand, this appears to rule out the
possibility that the [N {\sc ii}] and [S {\sc ii}] in the clump shadows is produced by
stellar UV, since we would expect the optical depths to 10-20 eV photons to be large within the clumps. 
Ionization behind the clumps may
be the result of UV photons released by Lyman-series recombinations;
however, such photons must originate in close proximity to the tails
in order to penetrate deeply into the shadow zone. Alternatively, these
regions may be exhibiting PDR-like behavior, wherein the layers
of [N {\sc ii}] and [S {\sc ii}] emission are the result of
significant N and S abundance enhancements driven by UV photodissociation of N-
and S-bearing molecules along the surfaces of the dusty clumps. Radio interferometric
imaging of such molecules (e.g., HCN, CN, SO, SO$_2$) in the clump regions would
test the latter hypothesis.

\subsection{The [Fe {\sc ii}] wedges}\label{sec:FeWedges}

In each lobe, the region (wedge) of bright [Fe {\sc ii}] lies on the opposite side of
the clump wedge from the amorphous wedge
(Fig.~\ref{fig:NGC6302zones}). The
1.64 $\mu$m [Fe {\sc ii}] emission line  has an
excitation temperature of $T_{\rm ex} \sim 1.1\times10^4$ K. As a consequence,
1.64 $\mu$m [Fe {\sc ii}] emission is typically a tracer of fast
(J-type) shocks. The line has been
detected and mapped in a diverse array of astrophysical environments,
such as Herbig-Haro objects associated with jets from young stellar
objects \citep[e.g.,][]{McCoey2004}, the Orion KL Nebula
\citep[e.g.,][]{Bally2015}, supernova remnants
\citep[e.g.,][]{Keohane2007}, and active galactic nuclei 
\citep[e.g.,][]{ForbesWard1993}; all of these systems feature shock speeds
in excess of $\sim$50 km s$^{-1}$ and often $>$100 km s$^{-1}$.
However, images of PNe in this
line are few and far between \citep[e.g., M2-9;][]{Balick2018}.  

The intensity of the 1.64 $\mu$m [Fe {\sc ii}] emission peaks sharply along the edges of these 
point-symmetric lobe structures (see N-S intensity profiles in lower panels of Fig.~\ref{fig:NGC6302ratioCuts}), 
indicating that these particular surfaces of the PN lobes are being actively
shaped by fast, collimated winds emanating from the immediate vicinity of
its central star. The lack of detection of X-rays from NGC 6302 by {\it Chandra} \citep{Kastner2012}
suggests that the relative wind speed at these [Fe {\sc ii}]-emitting interfaces cannot be much faster than necessary to excite the 1.64 $\mu$m line 
(i.e., $\sim$50--150 km s$^{-1}$). More sensitive soft X-ray imaging observations are clearly warranted, however,
especially given the detection of emission lines of O {\sc vii} and O {\sc viii} in archival IUE spectra \citep{Feibelman2001}.

\subsection{The ``amorphous'' wedges}

In contrast to the ``clump'' and [Fe II] zones just described, the ``amorphous'' wedge zones appear less structured, and display weaker 1.64 $\mu$m [Fe II] emission (Fig.~\ref{fig:NGC6302zones}).
Dynamically, these amorphous wedges may be ``quiet'' dynamic zones,
unlike the [Fe {\sc ii}] wedge regions (see next).  However, the azimuthal locations of the ``amorphous'' wedges closely 
coincide with faint, very extended, jet-like structures tentatively detected in wider-field UV imaging \citep{Rao2018}. 
The combination of thin-walled interfaces along the edges
of the amorphous zones and the possible UV detection of larger-scale structures 
along these same PAs \citep{Rao2018} indicates that fast winds
fill these apparently hollow (and seemingly dynamically quiet) lobe interiors and generate shocks along the inner lobe walls. 

Sensitive, high-resolution UV spectroscopy is required to confirm the apparent detection of
pc-scale, low-surface-brightness 
UV emission along the amorphous wedge directions by \citet{Rao2018} and, if confirmed, to ascertain whether this emission is in fact due to H$_2$ fluorescence, as proposed by those investigators. If this emission is indeed due to lines of H$_2$, such UV spectroscopic measurements might further yield the kinematic signatures of large-scale, collimated outflows.
Regardless, if fast winds are indeed flowing through the amorphous wedge
regions then --- given the symmetry of the amorphous and [Fe {\sc ii}]
wedges with respect to the central torus --- it stands to reason that these same winds cause the shocks that 
generate the bright, extended 1.64 $\mu$m emission within the [Fe {\sc ii}] wedges}. This then begs the
question as to why the 1.64 $\mu$m [Fe {\sc ii}] line is relatively faint and less extensive
within the amorphous wedges. The implication is either that the
density within the zone of fast wind flow within the amorphous wedges is
lower than that in the
[Fe {\sc ii}] wedges, or that the interactions of the
fast wind with the lobe walls are more energetic along the N (S)
borders of the W (E) lobes, where the strongest [Fe {\sc ii}] emission arises.

\subsection{The ``gap'' wedge}

The ``gap'' wedge is a narrow zone within the west lobe,
between the [Fe {\sc ii}] and clump wedges, that displays
a lack of structure. The gap wedge seems to be an empty, perhaps jet-evacuated region that
connects the inner west lobe to the larger ``breakout lobe'' structure to the NW (PA
$\sim$300$^\circ$) that is seen in wide-field images of NGC 6302 \citep[][]{Meaburn2005,Rao2018}. This
breakout lobe is expanding at an estimated deprojected
$\sim$600 km s$^{-1}$ in the 
plane of the sky at $\sim$7$’$ from
the nebular core \citep{Meaburn2005}. The breakout lobe is also seen to be
expanding laterally at $\sim$200  km s$^{-1}$ \citep{Lopez2012}.  This strongly implies that the same collimated
flow that generates the ``breakout lobe'' is also producing the
[Fe~{\sc ii}]-emitting shocks along the point-symmetric (ESE and WNW)
lobe surfaces. The ``gap'' wedge to the NW of the core region hence may simply be a
cowling through which invisible, supersonic, collimated winds flow into
the NW breakout lobe. This general
scenario has been validated via detailed hydrodynamic modeling of other bipolar pre-PNe and young PNe \citep{Balick2019}. 
Although both NW and SE breakout lobe structures are detected in far-UV imaging \citep{Rao2018}, 
the SE counterpart to the ``gap'' wedge is harder to discern in the WFC3 images (Fig.~\ref{fig:NGC6302zones}). 

\section{The central star: deepening enigma}\label{sec:CSPN}

\subsection{The imposter lurking near the core of NGC 6302}

\begin{figure}[!ht]
\centering
\includegraphics[width=7in]{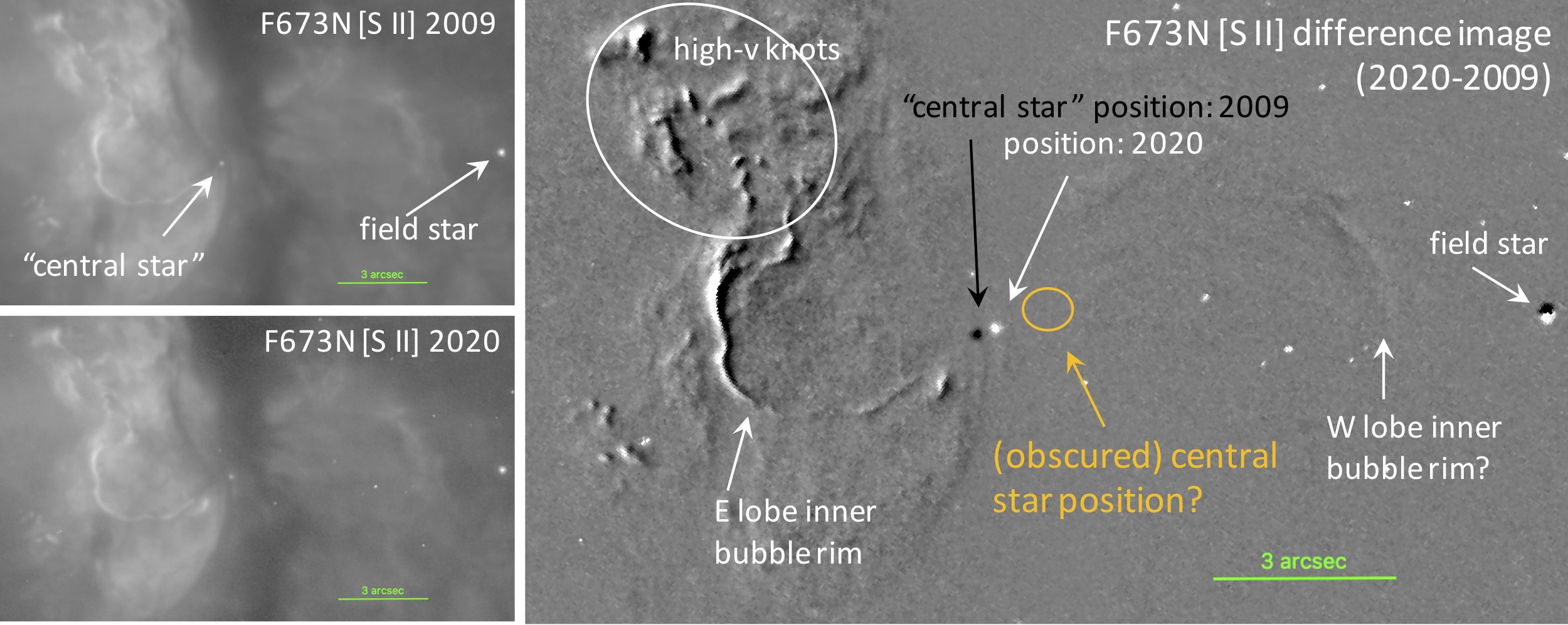}
\caption{
%Near-IR color montage of NGC 6302 (F160W: red, F130N: green, F110W: blue), with blowup (inset) illustrating 
Multi-epoch HST/WFC3 F673N imaging of the central $\sim17''\times10''$
region of NGC 6302. Left panels: F673N images of NGC 6302 obtained in 2009 (top) and in 2020 (bottom). Right panel: 2020$-$2009 F673N difference image. In each panel, N is up and E to left. The difference image establishes that the star previously identified as the CSPN \citep{Szyszka2009} exhibits proper motion of $\sim$35 mas yr$^{-1}$. 
Note the far smaller (and nearly orthogonal) proper motion of the field star at the far right of the frame.
The proposed actual position of the (obscured) central star (17:33:14.30, $-$37:06:12.244), as inferred from the inner lobe (high proper
motion, hence black/white) ``bubble'' features, is indicated with a yellow oval. A system of high-velocity knots (circled) extends to the NE of the inferred central star position. 
The uncorrected signatures of cosmic rays in the 2020 image appear as compact white spots in the difference image (with sizes less than those of the stellar images).
}
\label{fig:NGC6302centralStarLocation}
%\vspace{-.25in}
\end{figure}

As previously noted, we defer analysis of difference
images obtained from multi-epoch HST/WFC3 imaging to a future
paper. However, we report here on a striking result that is
immediately apparent from a difference image obtained from subtracting
the previous (2009) F673N image from our new
(2020) F673N image. The former WFC3 F673N image was used to identify a star, located near the intersection of the
nebula's waist and its symmetry axis, as the CSPN of NGC 6302
\citep{Szyszka2009}. The central region of the 
2009 and 2020 F673N images and the 2020$-$2009 F673N difference image is presented in
Fig.~\ref{fig:NGC6302centralStarLocation}. It is immediately apparent that the star that
was identified as the CSPN in the first-epoch (2009) WFC3 images in fact
shows significant proper motion (PM). 
Specifically, we measure a position of (RA, dec)
17:13:44.3632, $-$37:06:12.774 for this star at the epoch of the 2020
March 13 HST/WFC3 imaging, vs.\ 17:13:44.3883, $-$37:06:12.874 on 2009
July 27. This offset ($\sim$0.37$''$) translates to a PM of $\sim$35 mas yr$^{-1}$ at position
angle $\sim$285$^\circ$. This PM is larger than that measured for any other field star in the F673N image; all other 
field stars have PMs $<$10 mas yr$^{-1}$ (an example appears at the far right of the frame in Fig.~\ref{fig:NGC6302centralStarLocation}). 
Furthermore, these small field star PMs are directed randomly, as expected for accurate astrometric alignment. 
At $D = 1.0$ kpc, our adopted distance to NGC 6302, the PM  measured for the putative central star identified by \citet{Szyszka2009} would translate to a projected velocity of nearly 200 km s$^{-1}$  directed transverse to the equatorial plane of the system. This is a physically implausible velocity both for the CSPN of NGC 6302 or for a putative binary companion.

We conclude that the star previously identified as the central star of NGC
6302 is, in fact, unrelated to the nebula. Furthermore, while it is apparent from the prominence
of this star in our near-IR WFC3 images that it is quite
red, various lines of evidence indicate that this star is
most likely a foreground, as opposed to background, field
star  (the star's parallax cannot be
determined by Gaia, due to NGC 6302's bright core line
emission). From our 0.67 $\mu$m (F673N), 1.1 $\mu$m (F110W), and
1.6 $\mu$m  (F160W) images, we obtain magnitudes of 19.67$\pm$0.07, 16.74$\pm$0.02, and 16.15$\pm$0.02,
respectively. 
For distances between $\sim$250 pc and $\sim$1 kpc, the reddening along the line of sight to NGC 6302 is 
$E(B-V) \approx 0.3$ \citep{Lallement2019}, 
suggesting the dereddened F673N, F110W, and F160W magnitudes are approximately 18.9, 16.5, and 16.0, respectively \citep[][]{Cardelli1989}.
Assuming the field star is on the main sequence, these F673N$-$F110W and F110W$-$F160W colors (2.4 and 0.5 mag, respectively) 
indicate an early-M spectral type \citep{Pecaut2013}. We then (crudely) estimate a distance in the range $\sim$500--750 pc. The PM magnitude corresponds to a space 
velocity of $\sim$85 km s$^{-1}$ at 500 pc, suggesting that the
star may be a high-velocity (possibly Pop II) dwarf. While it is possible the star is instead
a cool giant that lies well behind the nebula, and is hence reddened by the dust in the
central torus region, this seems unlikely, given the 
space velocity would then be $\gtrsim$200 km s$^{-1}$. Furthermore, we
measure the same flux for the star (to within $\sim$10\%) in the two
F673N images, despite its (projected) position deeper within the dark lane in
the 2020 image. If this were a background star, one might have expected to 
observe a measureable change in brightness (most likely, a dimming).

\subsection{Implications for the properties of the central star and its progenitor}

It is truly an unfortunate coincidence that a foreground star displaying
the largest PM of any field star in the F673N image field of view
is seen projected so near the geometric center (waist)
of the bipolar nebula NGC 6302. Obviously it is necessary to discard
previous deductions concerning the properties of the central star of NGC 6302 that were
based, directly or indirectly, on the properties of this interloping
field star. In particular, conclusions concerning the
extinction toward the CSPN and the present-day CSPN mass that were
based on its apparent direct detection by HST
\citep[such as the $A_V$ estimates obtained
by][]{Szyszka2009,Wright2011} are certainly not
reliable. Fortunately, however,
most literature determinations of the central star's luminosity and effective
temperature --- including those in \citet{Szyszka2009} and
\citet{Wright2011} --- are based on the total luminosity and excitation of
the nebular emission, as opposed to the visible-wavelength
photospheric properties of the field star that, we now know, is merely seen in projection toward the
nebula's dark lane.

The newly established absence of the central star in HST imaging also
then begs the question: {\it Where, in fact, is the central star of
NGC 6302?} Obviously it is essential to answer this question, given
that all the available evidence points to the likelihood that this
object is descended from a star that pushed the upper envelope
of plausible PN progenitor masses.  In particular, the HST/WFC3 images confirm,
and further elaborate on, the findings of
\citet{Rauber2014} that the [N  {\sc ii}]/H$\alpha$ ratios of the lobe clumps are 
unusually high ([N  {\sc ii}]/H$\alpha$ $\sim$ 10; \S \ref{sec:lineRatios}). This supports previous abundance analyses
establishing that both N/H and N/O are very large in NGC 6302 \citep[i.e., super-solar by
factors of $\sim$10 and $\sim$200,
respectively;][]{Milingo2010,Rauber2014}, and strongly suggests that the 
clumps were ejected by a massive, N-enriched central AGB star
\citep{KarakasLugaro2016}.  This inference is consistent with the (large)
estimated luminosity and effective temperature of the present-day
central star \citep[i.e.,  $L_\star \gtrsim 4200 L_\odot$, adopting $D = 1.0$ kpc, and $T_{\rm
  eff} \sim 220$ kK;][]{Wright2011}. NGC 6302 is also among the most massive known PNe, with total gas mass estimates
ranging from $\sim$0.5 $M_\odot$ \citep{Dinh-V-Trung2008} to $\sim$3.5 $M_\odot$ \citep[][where this 
latter estimate has also been corrected for our adopted distance of $1.0$ kpc]{Wright2011}. 
%, $\sim$60\% of which is neutral \citep{Wright2011}. 
Estimates of the gas mass in the central torus region alone range from $\sim$0.1 $M_\odot$ \citep{Dinh-V-Trung2008,SantanderGarcia2017} to as high as $\sim$3 $M_\odot$ \citep[][but see discussion in \citealt{Dinh-V-Trung2008}]{Matsuura2005}.

%He/H		N/H		N/O		O/H		Ne/H		Ne/O
%0.156		4.25e-4		2.6		1.63e-4  	4.35e-5		0.267		KwitterHenry
%0.17  		7.06e-4		4.4		1.44e-4  	7.86e-5		0.546		Rauber
%0.086		0.63e-4		0.014		4.57e-4 	6.92e-5		0.15		solar

One potential clue to the location of the CSPN is offered by the F673N difference image
(Fig.~\ref{fig:NGC6302centralStarLocation}). Specifically, the east
lobe shows a rapidly expanding inner arc or rim of length $\sim$20$''$ that
is suggestive of the leading edge of a bubble-like structure. There appears to
be a faint counterpart to this expanding arc structure within the west lobe. 
If this pair of high-proper-motion, bubble-like features was produced by simultaneous, oppositely directed mass ejections, then the star could lie at their midpoint. As indicated in Fig.~\ref{fig:NGC6302centralStarLocation}, 
these inner shell features thus suggest the central star is actually
located well to the west of the foreground field star, right at the
heart of the nebula's central dark lane/torus, at a position of (RA,
dec) 17:33:44.30, $-$37:06:12.2 (where this estimate of the midpoint of the rims of the bubble-like features has an uncertainty of $\sim$0.5$''$). 
A system of high-PM knots, apparent in the F673N difference image (see Fig.~\ref{fig:NGC6302centralStarLocation}), 
appears to emanate from near this same position. No point source is detected at this
position in our HST/WFC3 images, however --- even in the near-IR (at wavelengths as long as
$\sim$1.6 $\mu$m) --- suggesting that $A_V >> 10$ \citep[see discussion in][]{Wright2011}. In future work, we will
test this predicted central star position, via determination of the expansion center of the
nebula as well as analysis of archival ALMA data
\citep{SantanderGarcia2017}. 

\subsection{What happened at the core of NGC 6302?}

All of the preceding lines of evidence --- NGC 6302's highly elevated N/H
abundance ratio, the large luminosity and effective temperature of
its central star, and its large inferred nebular mass --- support the hypothesis that the
present-day NGC 6302 nebula was generated by a star of initial mass $\sim$5--8
$M_\odot$. Such a star represents the upper
end of the progenitor mass range that can generate a PN, as opposed
to exploding as a supernova. Apparently the bulk of
the mass of this behemoth of a progenitor star was ejected over the
past $\sim$2000--5000 yr \citep{Meaburn2008,SantanderGarcia2017}, implying
a mass loss rate exceeding $10^{-4}$ $M_\odot$ yr$^{-1}$. This would far
surpass the mass loss rates typically observed for even the most
luminous AGB stars \citep{HofnerOlofsson2018}. Furthermore,
models predict that the evolutionary timescales of PN central stars as
massive, luminous, and hot as the CSPN of NGC 6302 are on the order of
centuries \citep[e.g.,][]{MillerBertolami2016}. It is
hence clear that --- as has been previously speculated by various
investigators over the past four decades \citep[e.g.,][]{Aller1981,Meaburn2008,Wright2011,Szyszka2011}
--- we are likely witnessing NGC 6302 at a key, highly transient evolutionary period immediately
following a highly disruptive event, or perhaps series of events, in the recent history of its central star system.
Use of the term ``central star system'' in this context is intentional (and essential) since, as noted in the Introduction, 
generation of pronounced bipolar structure, as observed in NGC 6302, most likely 
ultimately requires the presence of a close binary companion to the PN central star\footnote{Intriguingly, \citet{Feibelman2001} 
identified excess UV continuum and Mg {\sc ii} emission from the central region of NGC 6302, 
and commented that these emission sources were suggestive of the presence of a G-type companion to the central star.}.  

Combining the qualitative analysis presented in \S 5 
with the results of previous investigations of the structure and kinematics of NGC 6302, there is evidence that distinct episodes of such interacting-binary-influenced mass loss have led to the nebula we presently observe. Specifically, the dusty, molecule-rich torus that defines the equatorial plane of the system was evidently ejected, at low speed ($\sim$10 km s$^{-1}$), $\sim$5000--7500 years prior to the present observing epoch \citep{Peretto2007,SantanderGarcia2017}.  The N-rich clumps within the nebular lobes were then ejected in a dense but faster ($\sim$100 km s$^{-1}$) bipolar wind that was initiated some $\sim$3000 years after the ejection of the equatorial torus \citep{Szyszka2011}. The ejection of these lobe clump structures was accompanied or closely followed by much faster ($\sim$600 km s$^{-1}$) winds along the ``gap'' and (possibly) ``amorphous'' directions (see Fig.~\ref{fig:NGC6302zones}); the ($\sim$2000--2500 yr) dynamical ages of these far larger lobe/jet structures \citep{Meaburn2008}\footnote{\citet{Rao2018} deduce dynamical ages of $\sim$10 kyr for the UV-imaged lobe/jet structures on the basis of an assumed outflow velocity of 160 km s$^{-1}$, a factor $\sim$4 smaller than measured for the extended NW lobe structure of NGC 6302 \citep[][]{Meaburn2005}.} are similar to that of the ejection timescale of the lobe clumps. Our new HST/WFC3 imaging detection of extensive [Fe~{\sc ii}]-emitting shocks along the lobe walls and, to a lesser extent, within the ``clump'' and ``amorphous'' regions of the lobes has now revealed that this most recent stage of fast ($\stackrel{>}{\sim}$100 km s$^{-1}$) winds --- i.e., collimated and, possibly, precessing outflows from the putative binary system at NGC 6302's core region --- is ongoing. Finally, we note that,  according to models that simulate the formation of bipolar PNe, the foregoing episodes of directed mass loss likely would have been preceded by a sustained period of more or less isotropic AGB mass loss at lower rates \citep{LeeSahai2003,HuarteEspinosa2012,Balick2019}.
The extant data do not provide any direct evidence of this phase of quiescent mass loss, though they also do not rule out such a phase. 

Development of a self-consistent physical model of the rapid evolution of NGC 6302 that can account for all of these structural and kinematic elements is beyond the scope of this paper. We instead briefly consider two models, both involving interacting binary systems, that have been proposed to lead to abruptly transformative episodes during late (AGB) evolutionary stages of the primary star (the first of which having been proposed to apply specifically to NGC 6302, among other nebulae): (1) a so-called intermediate luminosity optical transient (ILOT) event, and (2) onset of common envelope evolution. 

\citet{SokerKashi2012} have hypothesized that NGC 6302 and a
handful of similarly extreme bipolar nebulae may have undergone 
ILOT events, during which much of the
mass in their bipolar lobes was ejected. Such ILOTs, which are intermediate in luminosity between novae and supernovae, have
been observed in a few extragalactic sources and luminous blue
variables (LBVs). \citet{SokerKashi2012} point out that the kinetic
energy represented by the lobes of NGC 6302 is
similar to the former class of eruptive object, whose members (collectively) lie along a ``stripe'' of energy-timescale
parameter space connecting them to LBVs. In relating NGC 6302 to
ILOT sources, they describe a scenario in
which an AGB star with a main-sequence companion in a (possibly
eccentric) few-au orbit undergoes short-lived, binary-interaction-driven phases of instability, during which
the AGB star is prone to lose mass at high
rates. 
\citet{SokerKashi2012} speculate that, prior to the ILOT episode, an interaction of the AGB star and companion (involving no Roche Lobe overflow) resulted in
the formation of the NGC 6302 equatorial torus. Subsequently --- a few $\times10^3$ yr after torus formation, given the  ``timeline'' of NGC~6302's mass loss episodes described above --- a second binary encounter, or series of encounters over a $\lesssim$100 yr period, led to Roche Lobe overflow. As a result, some fraction of the mass ejected during this second encounter was captured in an accretion disk around the companion, leading to the
formation of transient, accretion-powered jets. The resulting ILOT then generated the
bipolar lobes over this same relatively short period. 

If this ILOT-based model indeed describes the formation of NGC 6302, then it
would remain to explain the presence of the extensive ($\sim$0.3
pc-scale), point-symmetric regions of active shocks in the nebula that have now been 
revealed by our 1.64 $\mu$m [Fe {\sc ii}] imaging
(Fig~\ref{fig:NGC6302imageOverlay1}). That is, if these [Fe~{\sc ii}]-emitting shocks 
and the many other point-symmetric features within this nebula that are apparent in our imaging and in various
previous imaging studies of this nebula \citep[e.g.,][]{Szyszka2011,Rao2018} are generated by
(possibly precessing) jets from a binary companion's
accretion disk, then clearly the resulting wind interactions that have
shaped NGC 6302 are ongoing. Furthermore,  the 
region of NGC 6302 within $\sim$20$''$ of the central dark lane (faint circle
in Fig.~\ref{fig:NGC6302zones}) features a system of nested bubbles and knots with large PMs, oriented at a
range of PAs, that appear to emanate from the
presumed position of the CSPN (Fig.~\ref{fig:NGC6302centralStarLocation}). This (presumably) younger,
``multipolar'' core region somewhat resembles the complex cores of
the bipolar nebulae Hubble 5 and NGC 2440
\citep{CorradiSchwarz1993,LagoCosta2016}. All of the above features
of the nebula suggest that accretion processes are still active in 
the central binary system within NGC 6302. 

Alternatively, it is possible that NGC 6302 is the result of the onset
of a common envelope (CE) stage in the evolution of a close binary
system, a scenario also considered by \citet[][]{Rao2018}. Recent simulations have shown that such a CE event can result
in formation of a dense toroidal structure, which can in turn serve as
an effective collimator for a subsequent fast wind from the central, merged binary
remnant \citep{GarciaSegura2018,Zou2020}. This configuration results in simulated bipolar
nebulae with structural features (i.e., torus, lobes,
shocks) that overall resemble those observed in NGC 6302. 
However, as noted, the generation of large outer lobe structures and point-symmetry that characterize NGC 6302 likely requires pre-existing and ongoing jet activity, respectively, which in turn points to the influence of binary interactions prior to and following the common envelope phase \citep[see][]{GarciaSegura2018}.

\newpage

\section{Summary and Conclusions}

We have presented the results of a comprehensive, near-UV-to-near-IR Hubble Space Telescope WFC3 imaging
study of the young planetary nebula (PN) NGC~6302, the archetype of the class of
extreme bi-lobed, pinched-waist PNe that are rich in dust and
molecular gas. The new WFC3 emission-line image suite 
clearly defines the dusty toroidal equatorial structure that bisects NGC 6302's polar lobes, and the fine structures
(clumps, knots, and filaments) within the lobes. 
The most striking and unexpected aspect of the new WFC3
image suite is the bright, S-shaped 1.64 $\mu$m [Fe {\sc ii}] emission
that traces the southern interior of the east lobe rim 
and the northern interior of the west lobe rim, in point-symmetric
fashion, which we interpret as a zone of shocks caused by ongoing, fast
($\stackrel{\sim}{>}$100 km s$^{-1}$), collimated winds from NGC 6302's
central star(s). 

The [Fe {\sc ii}] emission zone represents one of a
handful of regions beyond $\sim$20$''$ from the central dark lane that
appear as distinct azimuthal zones (wedges) within the nebula. In
addition to the  [Fe {\sc ii}] zone, these  wedges comprise a region featuring
dense knots with outward facing features and tails along the nebular symmetry axis;
a dark wedge with traces of dust-scattered light along the nebular minor axis;
and  a zone with radial streaks or amorphous structure in between. The boundaries between these
wedge-shaped regions are marked by strong azimuthal
gradients in nebular ionization structure. 

These new (2019+2020) WFC3 images also reveal that the object previously
identified as NGC 6302's central star (on the basis of 2009 HST/WFC3
imaging)  is in fact a foreground field star. We propose that a pair
of bubble-like features in
the nebula's core region instead likely pinpoints the central star's actual
position within the nebula's dusty central torus. The available
evidence indicates that this star is descended from a progenitor near the upper
end of the mass range that could generate a PN, as opposed
to exploding as a supernova.

The features revealed by our panchromatic HST/WFC3 images of NGC 6302 --- in
particular, its distinct azimuthal structural zones and nested
bubble system, and the surprising
misalignment of the central engine's present collimated fast wind
direction (as traced by 1.64 $\mu$m [Fe {\sc ii}] emission) and the
nebula's main axis of symmetry (as defined by its dusty molecular
torus, polar-axis clump system, and outer lobe walls) --- presents an
especially daunting challenge for models of the origin and evolution
of bipolar structures in PNe. The presence of an interacting binary companion to the
central former AGB star, leading to formation of a disk/jet system,
appears to be a requirement of such a model. It may be impossible to establish
whether one star or two stars now lie at the core of NGC
6302, given the enormous extinction to and luminosity of
its central star system. Nonetheless, additional observations that can pinpoint the central
star(s) are clearly warranted. In particular, subarcsecond-resolution observations in the mid-IR (JWST) and submm (ALMA) should lead to
far more stringent constraints on both the position and energetics
of the central powering source of this enigmatic nebula.

\acknowledgements{Based on observations made with the NASA/ESA Hubble Space Telescope, obtained at the Space Telescope Science Institute (STScI), which is operated by the Association of Universities for Research in Astronomy, Inc., under NASA contract NAS5-26555. These observations are associated with program \#15953. Support for this program was provided by NASA through STScI grant HST-GO-15953.001-A to RIT. The authors wish to thank Jeremy Walsh, Stavros Akros, and the anonymous referee for comments that improved this manuscript.}

%Future work will involve bservational dynamics (proper motions and
%Doppler structure); evidence for and the roles of wind-induced
%shocks; hydro models that trace the histories of the various
%azimuthal structural zones. 



%\bibliography{PNeHSTsurveyRefs.bib}{}
%\bibliographystyle{aasjournal}

\end{document}